\documentclass[%
superscriptaddress,
preprint,
showpacs,preprintnumbers,
 amsmath,amssymb,
prb,
]{revtex4-1}

\usepackage{graphicx}
\usepackage{color}
\usepackage{dcolumn}
\usepackage{bm}
\usepackage{hyperref}

\begin{document}

\title{Magnetic Properties of Transition-Metal Adsorbed $ot$-Phosphorus Monolayer: A First-principles and Monte Carlo Study}



\author{Zengyao Wang}
\affiliation{Engineering Science Programme, Faculty of Engineering, National University of Singapore, Singapore 117575}
\author{Qingyun Wu}
\affiliation{Department of Materials Science Engineering, National University of Singapore, Singapore 117575}
\author{Lei Shen}
\email{shenlei@nus.edu.sg}
\affiliation{Department of Mechanical Engineering, Engineering Science Programme, National University of Singapore, Singapore 117575}

\date{\today}

\begin{abstract}

Using the first-principles and Monte Carlo methods, here we systematically study magnetic properties of monolayer octagonal-tetragonal phosphorus with 3$d$ transition-metal (TM) adatoms. Different from the puckered hexagonal black phosphorus monolayer (phosphorene or $\alpha$-P), the octagonal-tetragonal phase of 2D phosphorus (named as $ot$-P or $\varepsilon$-P in this article) is buckled with octagon-tetragon structure. Our calculations show that all TMs, except the closed-shell Zn atom, are able to strongly bind onto monolayer $ot$-P with significant binding energies. Local magnetic moments (up to 6 $\mu$B) on adatoms of Sc, Ti, V, Cr, Mn, Fe and Co originate from the exchange and crystal-field splitting of TM 3$d$ orbitals. The magnetic coupling between localized magnetic states of adatoms is dependent on adatomic distances and directions. Lastly, the uniformly magnetic order is investigated to screening two-dimensional dilute ferromagnets with high Curie temperature for applications of spintronics. It is found that $ot$-P with V atoms homogeneously adsorbed at the centre of octagons with a concentration of 5\% has the most stable ferromagnetic ground state. Its Curie temperature is estimated to be 173~K using the Monte Carlo method.
\end{abstract}

\maketitle

\section{Introduction}

The research scope of two-dimensional (2D) materials has developed from graphene \cite{butler2013progress} in the early stage into a vast range of layered materials
including transition metal dichalcogenies (TMDs) and different 2D
allotropes of silicon and phosphorus \cite{wang2012electronics, lopez2011boron, Reich2014Nature}. In particular, monolayer black phosphorus, known as phosphorene or $\alpha$-P, has drawn great attentions since its first successful fabrication in 2014 \cite{Li2014NN, Liu2014AN, Koenig2014APL}. Comparing to carbon-based 2D materials, the extra pair of electrons in a P atom allows the phosphorene monolayer to have a band gap which can be tuned by the strain, electric field or ``cutting" into ribbons \cite{Rodin2014PRL, Peng2014PRB, Zhu2014PRL, Zhu2015PRB, Guan2014PRB, Guan2016PRB, Fei2014NL, Fei2015PRB, Tran2014PRB, Tran2014PRB2, Ghosh2016PRB, Wu2015PRB, Gomes2015PRB, Ziletti2015PRB, Elahi2015PRB, Cakir2015PRB, Maity2016PRB, Ma2016PRB, Rivero2015PRB}. Furthermore, phosphorene shows good magnetic and/or magneto-optical properties \cite{Yang2016PRB, Seixas2015PRB, Hanakata2016PRB, Fu2015PRB, Tahir2015PRB, Jiang2015PRB, Pereira2015PRB, Ostahie2016PRB, Zhou2015PRB}. These properties make phosphorene a great potential in the applications of electronics, spintronics and photonics. The challenge of 2D black phosphorus for practical applications is its poor stability in the ambient environment, and thus it needs to be capsuled by other inert 2D materials, such as $h$-BN \cite{Koenig2014APL, Ziletti2015PRL, Ziletti2015PRB, Guan2014PRB, Rivero2015PRB}. Recently, some other layered allotropes of phosphorus beyond black phosphorus are proposed, such as blue phosphorus ($\beta$-P) \cite{Zhu2014PRL}, $\gamma$-P \cite{Guan2014PRL, Guan2014PRL2, Boulfelfel2012PRB} and $\delta$-P \cite{Guan2014PRL, Guan2014PRL2}. Very recently, a new octagonal-tetragonal phosphorus is emerged, which is composed of alternate octagon and tetragon (coined as $ot$-P or $\varepsilon$-P) instead of puckered hexagons in mono-layered black phosphorus as shown in \textbf{Fig. 1} \cite{Zhang2015CMS, Zhao2016CPL}. The pristine monolayer $ot$-P is stable and exhibits semiconducting properties [see \textbf{APPENDIX}]. As an elemental semiconductor $ot$-P allows its corresponding 2D type diluted magnetic semiconductors (DMSs) to have a more stable ferromagnetic state than III-V and II-VI based DMSs by preventing the formation of anti-site defects \cite{sato2010first, zhou2013intrinsic}. Furthermore, the anisotropic buckled structure of $ot$-P in principle allows for different coupling strength along different directions, which provides a way to tune the magnetic properties with doping. These properties suggest $ot$-P to be a promising material for 2D DMSs and spintronic applications.

In this work, a systematic study of the magnetic properties of transition-metal
(TM) atom (Sc-Zn) adsorbed $ot$-P monolayers ($ot$-P+TM) is carried out
by the first-principles and Monte-Carlo methods. It is
found that except for closed-shell atom Zn all of TM atoms are
firmly binding onto the $ot$-P monolayer with significant binding
energies. $ot$-P+TM systems for TM from Sc to Co exhibit local magnetic moments. The projected density of states shows that the localized magnetism mainly originates from the crystal field and exchange splitting of TM 3$d$ orbitals. The magnetic coupling and order acquired through the long-range interaction between diluted defects are investigated to screening
possible ferromagnets for spintronic applications. Lastly, the Curie temperature of ferromagnets is estimated by the Monte Carlo method. This article is organized as following:
The computational details are given in Sec. \textbf{II.} We present details of the optimized geometric structures and origin of local moments with an isolated adatom in Sec. \textbf{III} and subsections. The long-range magnetic coupling, magnetic order, and the Curie temperature of ferromagnetic systems is studied in Sec. \textbf{IV.} and subsections. At the end, the conclusions and discussions of this article are shown in Sec. \textbf{V} followed by the \textbf{APPENDIX} of the band structure and phonon dispersion.

\section{Computational details}

All the calculations are done via first-principles methods based on
density functional theory (DFT) \cite{hohenberg1964inhomogeneous, kohn1965self}, as implemented in the
\textit{Vienna Ab Initio Simulation Package} {\scriptsize{(VASP)}}\cite{Kresse1996PRB, Kresse1999PRB}. The exchange
correlation energy is simulated using generalized gradient approximation~(GGA)~in the form of the Perdew-Burke-Ernzerhof approximation (PBE) \cite{vasp2} functional, while the projector augmented wave (PAW) \cite{blochl1994projector}
approximation is used to describe the core electrons as external potentials
to the orbitals of study. For structure relaxation and calculation
of the binding energies, a supercell of $ot$-P lattice with centered
sampling point in the first Brillouin zone is used. In the perpendicular
direction to the monolayer, a vacuum layer of 30 \AA~is used to eliminate
interaction between adjacent slabs. The lattice constants for the
primitive unit cell of pristine $ot$-P monolayer used in this work are
$a=b=$6.455~\AA. the k-points mesh used for calculation
carried out in a $2\times2$ supercell is $3\times3\times1$. When
studying the influence of adsorption concentration on Curie Temperature,
systems with 5 different adsorption concentrations, i.e. $3.125\%,$
$5\%,$ $6.25\%,$ $12.5\%,$ and $25\%$ are studied, in which the
adsorption concentration is defined to be the ratio of TM atoms vs
P atoms in the system. Spin polarized calculations are performed throughout
the work. The kinetic energy cutoff for the plane wave basis set is
chosen to be 255~eV, which yields well-converged total energies.
All the structures are relaxed until the remaining force on each atom
is reduced to less than 0.01~eV/\AA.

\begin{figure*}[ptb]
\centering
\includegraphics [width=0.6\textwidth]{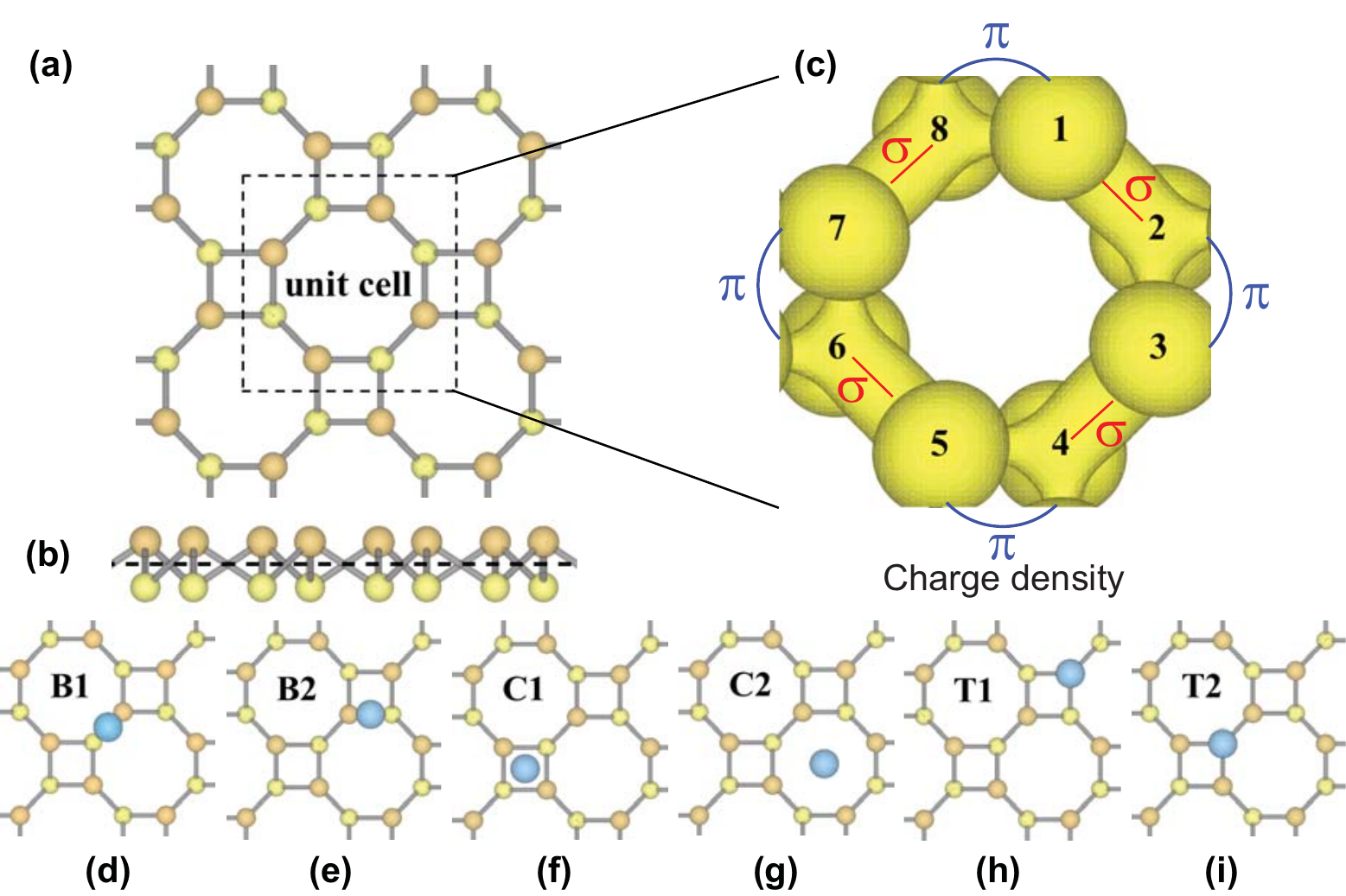}
\caption{{\footnotesize{}Front view (a) and side view (b) of a 2D $ot$-P monolayer. The unit cell of $ot$-P consists of 8 atoms, which is denoted by a square in (a). The $ot$-P monolayer has a non-planar buckled structure, with half of the eight atoms at one side of a hypothetical middle plane and half at the other side. (c) Charge density distribution at the $\Gamma$ point of the valence band of 2D $ot$-P. The atoms labeled with 1, 3, 5, and 7 are in front of the hypothetic middle plane, while atoms labeled 2, 4, 6 and 8 are behind the middle plane. Each atom forms one $\pi$ and one $\sigma$ bonding with its two nearest-neighbor atoms through the occupied $p$ orbital. For clarity, in (a)(b) and all the following figures, we use orange color (dark grey) to represent atoms above the middle plane and yellow color (light grey) to represent atoms below the middle plane. (d)-(i) $ot$-P monolayer with adsorption of a TM atom at $B_{1},$ $B_{2},$ $C_{1}$ $C_{2},$ $T_{1},$ and $T_{2}$ sites respectively. Our calculation shows that $C_{2}$ and $T_{2}$ positions are two stable adsorption sites among the six configurations. \label{fig:$ot$-P+AdSites}}}
\label{fig1}
\end{figure*}

\section{Isolated adatom}

First of all, we study geometric and electronic structures of a supercell of $ot$-P with an isolated TM adatom by placing a TM atom on a big supercell. The possible site of adsorption and local structural deformation after adsorption is investigated for all 3$d$ transition metals on the $ot$-P supercell. The origin of the localized magnetic state induced by the adatom is discussed in detail.

\subsection{Stability of geometric structures}

\begin{figure*}[ptb]
\centering
\includegraphics [width=0.75\textwidth]{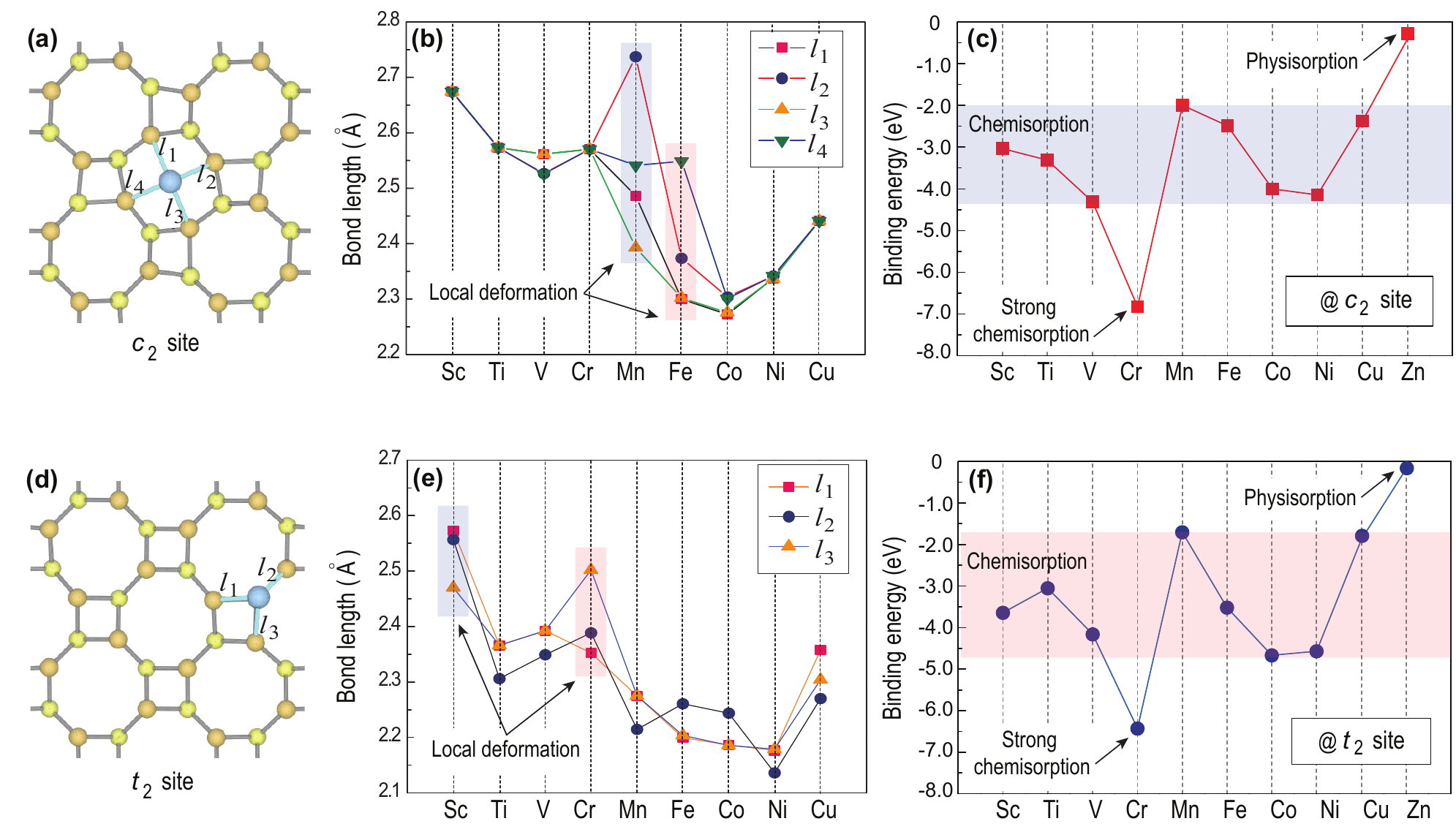}
\caption{{\footnotesize{}Optimized geometry of (a)TM@$C_{2}$ and
(c) TM@$T_{2}$ systems. The length of the TM-P bonds in
(b) TM@$C_{2}$ systems and (d)TM@$T_{2}$ systems for TM from
Sc to Cu. Notice that the Zn atom does not form bonds with nearby P atoms due to
its closed-shell structure. Before adsorbing, $l_1$, $l_2$, $l_3$ and $l_4$ are equivalent in (a), and $l_1$/$l_3$ (not $l_2$) are equivalent in (c). After adsorbing, they may not be equivalent due to the local deformation of the structure. The binding energies $(E_{b})$ of the TM atoms on
$ot$-P monolayers. Most of the 3$d$ TM atoms, except for closed-shell Zn, have a chemisorption on the $ot$-P monolayer with a large binding energy ($>$1.5~eV).\label{fig:OptimizedGeometry+BondLength}}}
\end{figure*}

Before we study the magnetic property of $ot$-P monolayers with an isolated TM adatom, we first check their stability of the optimized geometric structure. Different from planar structure of graphene or $h$-BN, the $ot$-P monolayer has a non-planar buckled structure. A unit cell of $ot$-P consists of eight phosphorus atoms, denoted by a square in \textbf{Fig. 1(a)}. In a unit cell, half of eight atoms are at one side of the hypothetical middle plane, while the other half are at the opposite side, as shown in \textbf{Figs. 1(b)} and \textbf{1(c)}. \textbf{Figure 1(c)} shows the charge density at the $\Gamma$ point of the valence band. Atoms labeled with 1, 3, 5 and 7 are in front of the hypothetic middle plane, while atoms 2, 4, 6, and 8 are behind the middle plane. As can be seen, there is only occupied bonding orbital (no anti-bonding orbital) in buckled $ot$-P. Each atom forms one $\pi$ and one $\sigma$ bonding with its two nearest-neighbor atoms through the occupied $p$ orbital. Even in the graphene structure, there is no strong short-range $\pi$ bonding between two adjacent carbon atoms, instead of a ``big $\pi$" bond floating over two sides of the graphene sheet. This all-bonding nature indicates an extremely stable structure of buckled $ot$-P, which is further verified by the phonon dispersion [\textbf{APPENDIX}].

 For all $ot$-P+TM
systems (TM = Sc, Ti, V, Cr, Mn, Fe, Co, Ni, Cu, Zn), six possible
adsorption sites are investigated, the bridge site over a P-P bond
at the side of an octagonal ring ($B_{1}$), the bridge site over
a P-P bond at the side of a tetragonal ring ($B_{2}$), the center
of a tetragonal ring ($C_{1}$), the center of an octagonal ring ($C_{2}$),
the top site of a P atom above the middle plain ($T_{1}$), and the
top site of a P atom below the middle plain ($T_{2}$), as shown in
\textbf{Figs 1(d)-(i)}. Through the optimization of the geometric structures of $ot$-P+TMs, we find that the more preferred sites are $C_{2}$ and $T_{2}$ positions. Other four sites of adatoms are found to be less energetically favourable. Regardless of their initial positions, the adatoms will spontaneously move to either $C_{2}$ or $T_{2}$ position after structural relaxation.
Therefore, in the following studies only systems of $ot$-P+TM atoms adsorbed on $C_{2}$ and $T_{2}$ sites are discussed.

Except for closed shell atom Zn, other 3$d$ TM atoms at either $C_{2}$ or $T_{2}$
site is able to bond covalently with nearby P atoms. For adsorption
at $C_{2}$ positions, the four covalent bonds are formed between
the adatoms and the four P atoms of an octagonal ring above the middle
plain. For $T_{2}$ site, adatoms are covalently bond to the three
nearest P atoms, which together with the other P atom below form a
valley at the joint corner of the octagonal and tetragonal rings.
Graphical representation of the covalent bonds and their lengths are
shown in \textbf{Fig. 2}. Different values of the bond length indicate that the adsorption of TM atoms may induce a local structural deformation. Mn and Fe can induce a large deformation at the $C_2$ site, while Cr makes a distinguished deformation at the $T_2$ site. Different values of bond length also mean different binding energies, $E_{b}$, which is defined as
\begin{equation}
E_{b}=E_{TM+ot-P}-E_{ot-P}-E_{TM},\label{Eb}
\end{equation}
where $E_{ot-P}$, $E_{TM}$ and $E_{TM+ot-P}$ are the total energies
of the pristine $ot$-P monolayer, the isolated TM atom, and the system
of $ot$-P monolayer with TM atoms adsorbed respectively. The calculated binding energies are shown in \textbf{Figs. 2(c)} and \textbf{2(f)}. It shows that except for the
closed shell transition metal atom Zn that is physisorbed to $ot$-P monolayer
with a binding energy as low as 0.17$\sim$0.40~eV, other transition
metal atoms are able to bond strongly to the monolayer with their
binding energies ranging from 1.37~eV to 3.98~eV. The proposed
structures are thus believed to be stable given their high binding
energies.

\begin{figure*}[ptb]
\centering
\includegraphics [width=0.75\textwidth]{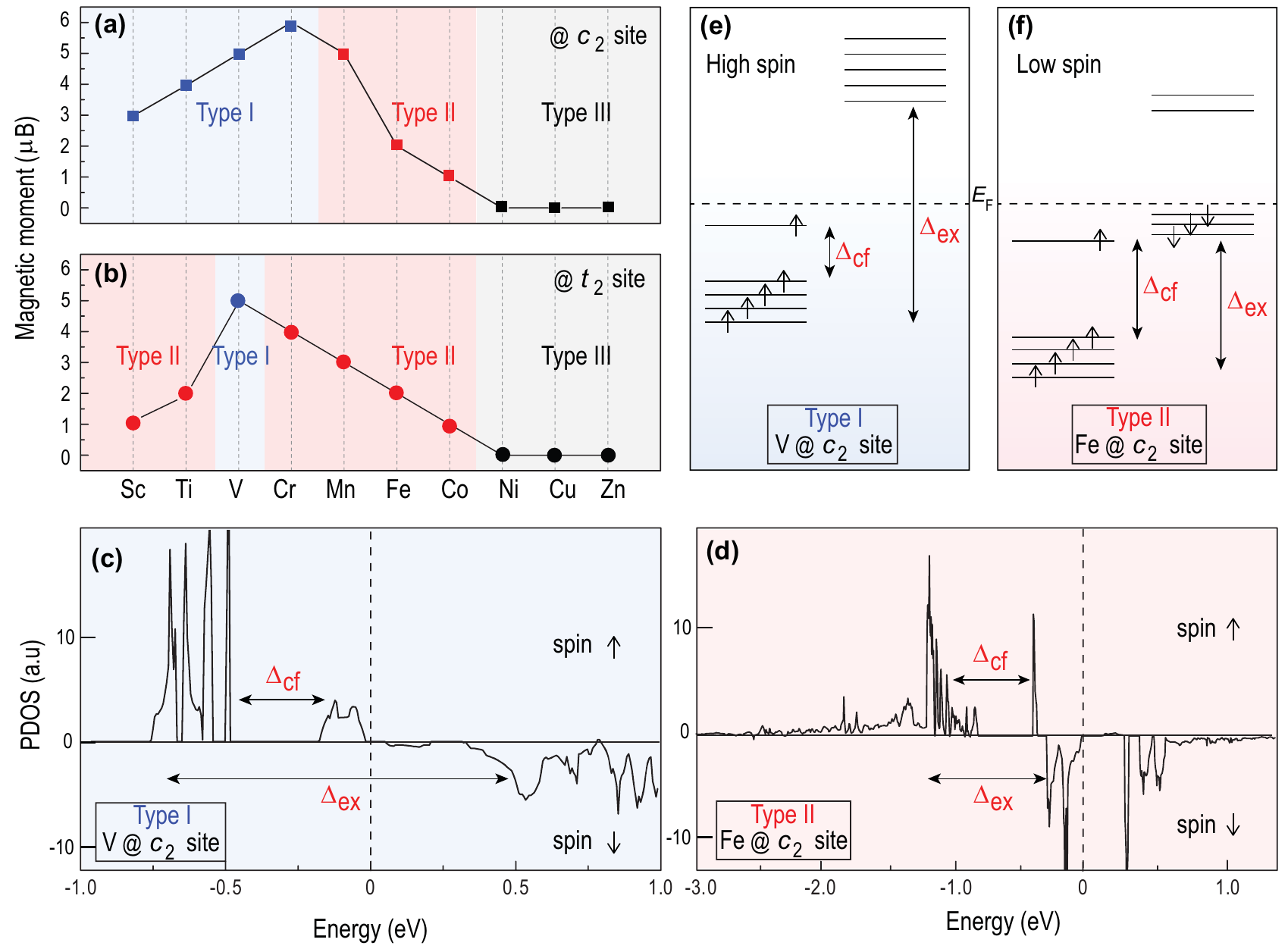}
\caption{{\footnotesize{}Magnetic moments of (a) $ot$-P+TM@$C_{2}$ and (b) $ot$-P+TM@$T_{2}$ systems. Each kind of systems is further divided into three types, Type I (high spin state), Type II (low spin state), and Type III (non-spin state). Partial density of states of (c) $ot$-P+V@$C_{2}$ and (d) $ot$-P+Fe@$C_{2}$ systems. Schematics of 3$d$ energy levels of (e) V (type I) and (f) Fe (type II) atom after adsorption on the $C_{2}$ sites of $ot$-P monolayer. In $ot$-P+V@$C_{2}$, the type I system, the crystal field splitting $\Delta_{cf}$ is smaller than the exchange splitting $\Delta_{ex}$, leading a high spin state and high total magnetic moment. On the other hand, in $ot$-P+Fe@$C_{2}$, the type II system, $\Delta_{ex}$ is comparable to $\Delta_{cf}$, resulting a low spin state and low total magnetic moment of $2\mu B$ due to the spin-moment compensation. \label{fig:MagMom+PDOS+ELevel}}}
\end{figure*}

\subsection{Origin and classification of localized magnetic states}

The magnetic moments of different $ot$-P+TM systems are shown in \textbf{Figs. 3(a)} and \textbf{3(b)}. The $ot$-P+TM systems are magnetic for TM atoms from Sc to Co for adsorption at both $C_{2}$ and $T_{2}$ sites. For transition metal atoms of Ni, Cu and Zn, the $ot$-P+TM systems are nonmagnetic regardless of their adsorption sites.

It can be seen in \textbf{Fig. 3(a)} that the highest magnetic moment
for $ot$-P+TM systems can reach $6\mu B$ per adatom due to the extra contribution
from 4$s$ electrons. In an isolated TM atom, the
4$s$ orbitals are of lower energies than the 3$d$ orbitals. However, after
adsorption on pristine $ot$-P monolayer, the large coulomb repulsion
between the monolayer and the 4$s$ orbitals of the TM atoms would push
up the 4$s$ orbitals. Electrons, originally in 4$s$ orbitals, would
instead occupy lower 3$d$ orbitals. These electrons would therefore also
contribute to magnetism of the systems. $ot$-P+TM systems can be further
classified into three types for each adsorption sites, the high spin
state (type I), the low spin state (type II), and non-spin state (type III). The classification
is made basing on the degree of crystal field splitting $\Delta_{cf}$
and exchange splitting $\Delta_{ex}$ of 3$d$ orbitals in different $ot$-P+TM systems.

The effect of crystal field splitting $\Delta_{cf}$ originates from
symmetry breaking when introducing the TM atoms to the $ot$-P monolayer
system. $ot$-P+TM systems with adsorption on $C_{2}$ sites are of $D_{4}$
symmetry, i.e. there is a four-fold axis perpendicular to the hypothetical
middle plane of $ot$-P and four two-fold axis normal to the four-fold
axis. On the other hand, $ot$-P+TM systems with adsorption on $T_{2}$
sites are of $C_{s}$ symmetry, which is lower than the $D_{4}$ symmetry. The crystal
field splitting effect in such systems is therefore more pronounced,
resulting in a lower degeneracy of 3$d$ orbital energy levels. Comparing to the PDOS
of V@$C_{2}$, there are more well-spaced discrete peaks
in the PDOS of V@$T_{2}$(not shown here), suggesting a lower degeneracy
of energy levels in the V@$T_{2}$ system. In the meantime, different TM atoms would also induce different extend of lattice distortion at the adsorption sites, which further breakdown
the symmetry of the system. For this reason, different TM atoms adsorbed
at the same site also have different degrees of crystal field splitting
of their 3d orbital energy levels. The relative values of the crystal
field splitting of the system to their exchange field splitting of
3$d$ orbitals under external magnetic field is the essential criterion
for determination of the high spin and low spin type of the system.

The exchange splitting $\Delta_{ex}$ is the energy difference of two electronic states due to exchange interactions, i.e., direct overlap of electronic wavefunctions. When the crystal field splitting $\Delta_{cf}$ is less pronounced than the exchange splitting $\Delta_{ex}$, the energy levels are
of higher degeneracy, as shown in \textbf{Figs. 3(c) }and \textbf{3(e)}.
the schematic representation of energy levels of V@$C_{2}$.
It is easier for the system to occupy all the spin-up states first
before starting to occupy the spin-down states. Thus, such systems
are expected to be able to reach higher total magnetic moment. This type of systems is of high
spin case and is classified as type I in this work. On the contrary,
when the crystal field splitting $\Delta_{cf}$ is more pronounced
than the exchange field splitting $\Delta_{ex}$, such system has a small amount of magnetic moments, as
shown in \textbf{Figs. 3(d) }and \textbf{3(f)}. It is because of the occupation of the spin-down states of the lower energy levels, thus cancels the total spin of electrons and results in a
lower total magnetic moment. For this reason, this type of systems
is classified as type II \textendash{} the low spin type. When the the exchange field splitting $\Delta_{ex}$ is zero, for example the absorptions of Ni, Cu and Zn atoms on $ot$-P monolayers at both $C_2$ and $T_2$ sites, these systems are nonmagnetic, defined as type III.

Using the above definition for type I (high spin), type II (low
spin), and type III (non-spin), we can classify all cases of $ot$-P+TM at both $C_2$ and $T_2$ sites. For systems with adsorption
at $C_{2}$ sites, systems with TM = Sc, Ti, V and Cr are of Type
I, and systems with TM = Mn, Fe and Co are of type II [\textbf{Fig. 3(a)}]. For systems with
adsorption at $T_2$ sites, systems with TM = Sc, Ti, Cr, Mn, Fe and Co are of type II
and the system with TM = V is of type I [\textbf{Fig. 3(b)}]. The all rest are of type III [\textbf{Fig. 3(a)} and \textbf{Fig. 3(b)}]. In particular,
the highest magnetic moment in TM@$C_{2}$ systems is $6\ensuremath{\mu}B$
per adatom, whereas the highest magnetic moment in TM@$T_{2}$
systems is $5\ensuremath{\mu}B$ per adatom. This is because for adsorption
at $C_{2}$ sites there is higher symmetry and less splitting of 3$d$
orbital energy levels than that at $T_{2}$ sites. Meanwhile, the exchange field splitting effect is large enough to completely separate the spin-up and spin-down states of
in total six 3$d$+4$s$ orbitals, resulting in a high total magnetic moment. Our PDOS results show that for Ni+$ot$-P systems the ten outmost electrons exactly fill up the
3d orbitals, thus having a spin configuration of no total magnetic
moment. For $ot$-P+Cu systems, the left one electron after filling up
the spin-up and spin-down states of 3d orbitals has a delocalized
nature and does not contribute to the local magnetic moment. As for
Zn, the close-shell atom is physisorbed onto the $ot$-P monolayer, thus
is able to maintain its atomic state without magnetism.

\section{Magnetic coupling, order and Curie temperature}

\begin{figure}[ptb]
\centering
\includegraphics [width=0.75\textwidth]{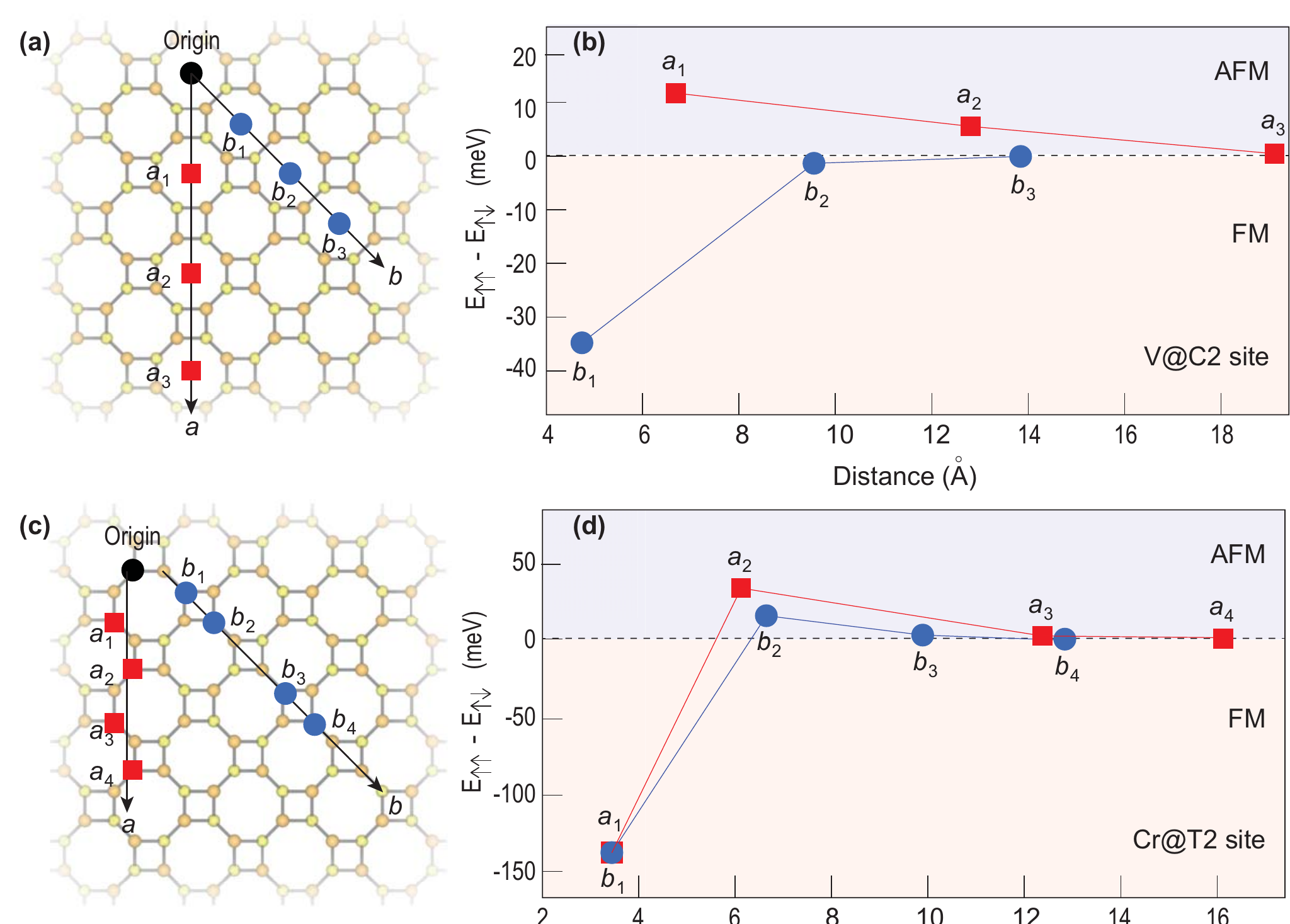}
\caption{{\footnotesize{}(a) Schematic representation of $C_2$-type sites along two different directions. The $a$ direction is defined as the one along which there are alternative tetragon and octagon rings, whereas along the $b$ direction the octagon rings are directly connected by sharing one side with each adjacent octagon ring. (b) schematic representation of $T_2$-type sites along two different directions. Energy differences between parallel and antiparallel spins as a function of distance and direction for (c) V@$C_2$ and (d) Cr@$T_2$. \label{FM-AFM-Ediff}}}
\end{figure}

After studying the localized magnetic state of $ot$-P with a single adatom, we next investigate the long-range magnetic coupling and order under different concentrations and directions of adatoms. After getting configurations with ferromagnetic order, the Curie temperature of ferromagnets is estimated and discussed for the application of spintronics. In this section, V and Cr are chosen as examples to study the effect of adsorption distance, direction, and concentration on magnetic properties of $ot$-P+TM systems. They are chosen over other TM atoms because these two elements exhibit
a high magnetic moment and strong binding energy at both $C_{2}$ and $T_{2}$ adsorption
sites as what has been shown in the previous section.

\subsection{Magnetic coupling}

In order to study the long-range magnetic coupling between two adatoms, we model a supercell of a $ot$-P monolayer with two adatoms at different adsorption sites and along different directions. One adatom is fixed as the origin, and then the other adatom is placed away along two different directions [\textbf{Fig. 4}]. Next, we study the long-range magnetic interaction between the origin and the other TM atom (V@$C_2$ and Cr@$T_2$ as examples) by calculating the energy difference between parallel- ($E^{\uparrow\uparrow}$) and antiparallel-spin ($E^{\uparrow\downarrow}$) configurations. Anisotropic (isotropic) magnetic interaction is found for adsorption at $C_2$ ($T_2$) sites. At the $C_2$ site, the exchange interaction aligns the spin antiparallel along the $a$ direction, while aligns the spin parallel along the $b$ direction [\textbf{Figs. 4(a)} and \textbf{4(b)}]. As for $T_2$ sites, along both $a$ and $b$ directions the exchange interaction aligns the spin parallel for short distances and spin antiparallel for long distances as shown in \textbf{Figs. 4(c)} and \textbf{4(d)}. There is no magnetic coupling ($\Delta_{E^{\uparrow\uparrow}-E^{\uparrow\downarrow}}=0$) when the separation of two adatoms is larger than 13~\AA.
\begin{figure*}[ptb]
\centering
\includegraphics [width=0.7\textwidth]{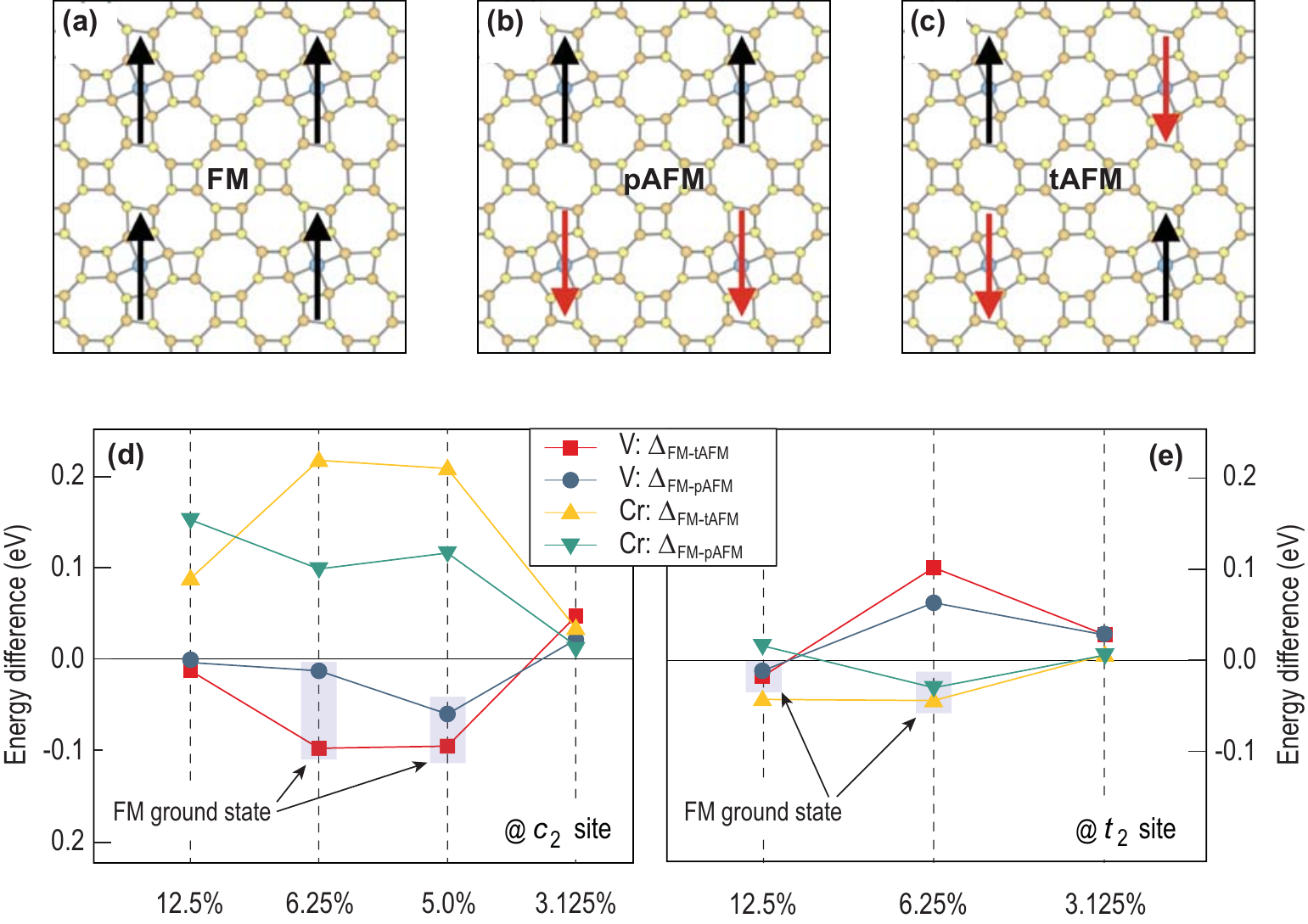}
\caption{{\footnotesize{}(a)-(c) The ferromagnetic (FM), parallel antiferromagnetic
(pAFM) and total antiferromagnetic (tAFM) order of the
TM adatoms in a $4\times4$ supercell at the $C_2$ site, corresponding to an adsorption
concentration of $3.125\%$ (Other concentrations and $T_2$-site cases are not shown here). Energy difference between FM and pAFM orders, FM and tAFM orders as function of adsorption concentrations
for (d) V/Cr@$C_2$ and (e) V/Cr@$T_2$ systems.\label{FM-AFM-Ediff}}}
\end{figure*}

\subsection{Magnetic order}

In order to study the magnetic order under different concentrations, we
use a supercell with 4 adsorption atoms, which are evenly distributed along both horizontal and vertical directions as shown in \textbf{Figs. 5(a)-(c)}. Different adsorption concentrations are controlled by adjusting the size of the supercell while keep the number of adatoms in each supercell
unchanged. Given two types of adatoms and two different adsorption sites, there are in total four categories to be considered, V atoms adsorbed at $C_{2}$ and $T_{2}$ sites, and Cr atoms adsorbed at $C_{2}$ and $T_{2}$ sites. For adsorption at $C_{2}$ sites, four cases corresponding to the adsorption concentration of $3.125\%$, $5\%$, $6.25\%$, and $12.5\%$ are considered. On the other hand, as $T_{2}$
sites are of lower symmetry than $C_{2}$ sites, isotropic systems are only able to be constructed with three adsorption concentration of $3.125\%$,$6.25\%$ and $12.5\%.$

To determine the magnetic order of the ground state, the energy difference of
systems with different magnetic orders are calculated as shown in \textbf{Figs. 5(d)} and \textbf{5(e)}. In this work, we considered two types of antiferromagnetic spin configurations, as shown in \textbf{Figs. 5(b)} and \textbf{5(c)}. In particular, the parallel antiferromagnetic configuration has one row of adatoms
with the same sign of spin and the adjacent two rows of adatoms
having spin of the opposite sign [\textbf{Fig. 5(b)}]. On the other hand, any
adatom in the total antiferromagnetic configuration has opposite sign
of spin with its four nearest neighbors[\textbf{Fig. 5(c)}]. Total energy of FM, pAFM and tAFM configurations is calculated and compared. It is found that V@$C_{2}$ systems with $5\%$ and $6.25\%$, V@$T_{2}$ system with $12.5\%$ and Cr@$T_{2}$ system with $6.25\%$ adsorption concentration are ferromagnetic. Next we explore the magnetic stability of these five ferromagnets by estimating their Curie temperature using the Monte Carlo method.

\subsection{Curie temperature of ferromagnets}

As $ot$-P monolayer is of square lattice geometry, the Curie Temperature
of the system can be calculated basing on 2D square lattice Ising
model. Considering nearest and second-nearest-neighbor interactions
without external field, the Hamiltonian of the 2D square Ising model
is
\begin{equation}
H(\sigma)=-J\sum_{<i,j>}\sigma_{i}\sigma_{j}-K\sum_{,<m,n>}\sigma_{m}\sigma_{n},\label{eq:ising model}
\end{equation}
where $J$ and $K$ are the nearest and second nearest coupling constant
respectively, and $\sigma_{i},$ $\sigma_{j},$ $\sigma_{m}$ and
$\sigma_{n}$ are discrete variables representing the site's spin
that belong to $\{+1,-1\}$. In this equation, the first sum is over
pairs of adjacent spins, and the second sum is over pairs of second-nearest
neighbors. The coupling strengths between the nearest sites and the
second nearest sites can be obtained from the energy difference between
systems with different spin configurations. From the Ising model, we can
write the total energy of ferromagnetic, total antiferromagnetic and
parallel antiferromagnetic systems as follows:

\begin{figure}[ptb]
\centering
\includegraphics [width=0.6\textwidth]{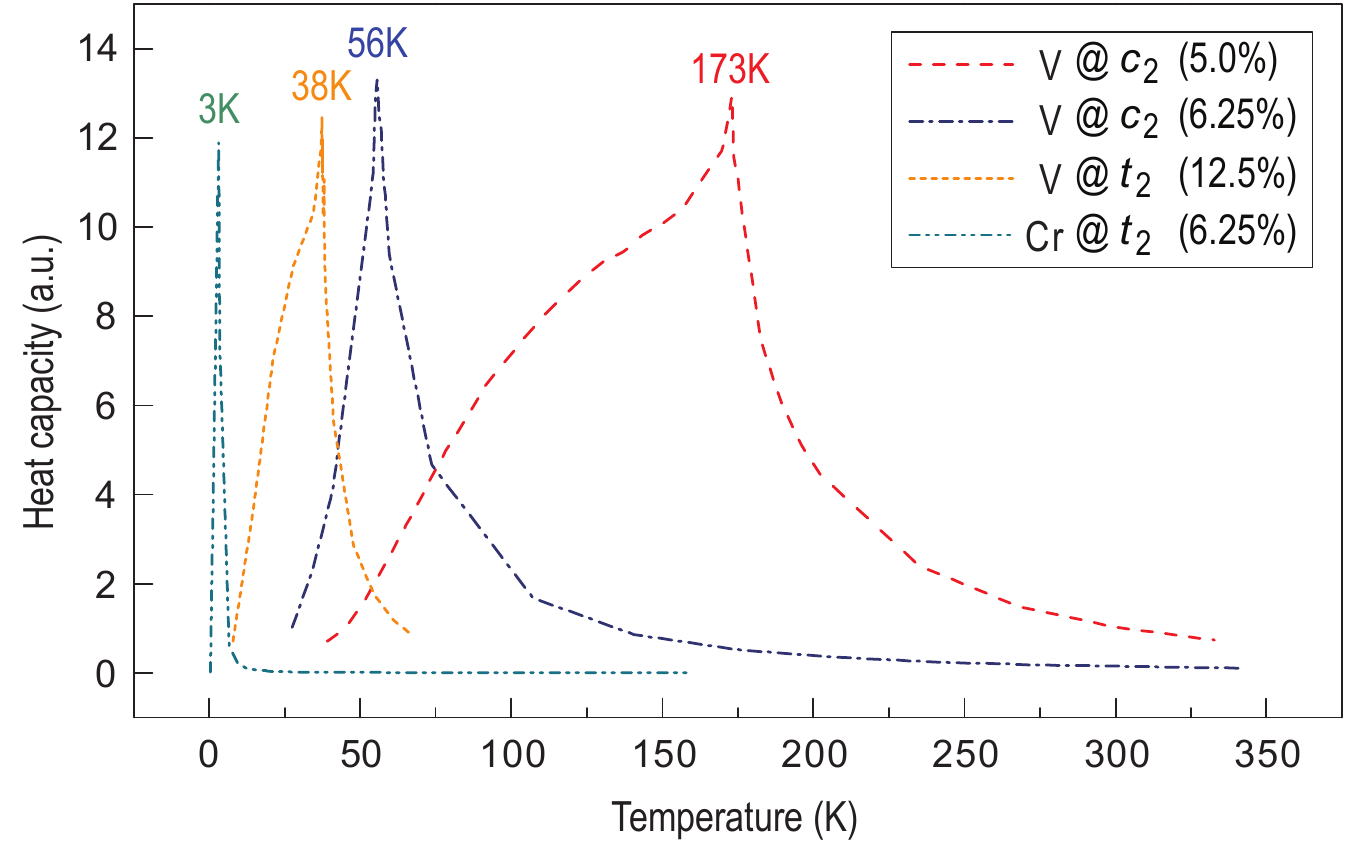}
\caption{{\footnotesize{}The temperature-dependent heat capacity $(C)$ of
all the previously found ferromagnetic systems. The peak of the heat
capacity $(C)$ to temperature $T$ curve corresponding to the ferromagnetic-to-paramagnetic
phase transition temperature, i.e. the Curie temperature. Among all
the systems, $ot$-P+V@C2 system with adsorption concentration of $5\%$
is found to exhibit the highest Curie temperature of 173K. \label{fig:T-dependent-C}}}
\end{figure}

\begin{equation}
E_{FM}=E_{nm}-8J-8K,\label{eq:EFM}
\end{equation}
\begin{equation}
E_{tAFM}=E_{nm}+8J-8K,\label{eq:EtAFM}
\end{equation}
\begin{equation}
E_{pAFM}=E_{nm}+8K.\label{eq:EpAFM}
\end{equation}
By adding and subtracting equation \ref{eq:EFM}, \ref{eq:EtAFM}
and \ref{eq:EpAFM}, we can obtain the expression for $J$ and $K$
as:
\begin{equation}
J=\frac{E_{tAFM}-E_{FM}}{16},\label{eq:J}
\end{equation}

\begin{equation}
K=-\frac{E_{tAFM}+E_{FM}-E_{pAFM}}{32}.\label{eq:K}
\end{equation}
The Curie Temperature is the temperature at which the ferromagnetic-paramagnetic
transition happens. It is expected that the peak of the temperature-dependent
heat capacity $(C)$ of the system will be observed at phase transition
temperature, i.e. the Curie Temperature. Here, the heat capacity $(C)$
of the system can be calculated using the expression:
\begin{equation}
C=\lim_{\Delta T\rightarrow0}\frac{\Delta E_{T}}{\Delta T},\label{eq:HeatCapacity}
\end{equation}
where $\Delta E_{T}$ is the change of the total energy of the system
caused by redistribution of spins when the temperature is increased
by $\Delta T$.

In this work, we used $100\times100$ lattice points containing $10^{4}$
local magnetic moments to sample our systems. For each system at each
temperature the simulation of spin distribution is looped for $2\times10^{8}$
times. By observing the peak of the temperature-dependent heat capacity
of all the systems, we found that V@$C_{2}$ system with adsorption
concentration of $5\%$ have the highest Curie temperature of 173~K as shown in \textbf{Fig. 6}. Such low adsorption concentration makes it possible in the experiment \cite{Dietl2010NM, Pan2007PRL, Sato2003EPL}. Even though the 173~K Curie temperature is still lower than the room temperature, it is larger than most traditional DMS, such as $5\%$ Mn doped GaAs ($T_c \sim 120$~K) \cite{Dietl2010NM}, $5\%$ C doped ZnO ($T_c = 80$~K) \cite{Pan2007PRL}. Due to the quantum confinement effect, the Curie temperature of zigzag phosphorene nanoribbons is estimated to be above room temperature using the Monte Carlo method \cite{Yang2016PRB}. Thus, the $ot$-P nanoribbons \cite{Zhao2016CPL} may have higher $T_c$ than 173~K and above room temperature. Notice that the mean-field approximation usually overestimates the Curie temperature \cite{Sato2003EPL,Seixas2015PRB}. For example, the Curie temperature of the same $ot$-P with $5\%~$V@$C_{2}$ is around 600~K if we use the mean-field approximation.\\ 

Using the first-principles calculations, we find that $ot$-P+TM systems have high structural
stabilities and able to exhibit collective magnetic moments for adatoms
from Sc to Co. The adatoms with localized magnetism on $ot$-P monolayers provide magnetic active sites, which can be used in sensors for detecting magnetic gas molecules. In particular, V@$C_{2}$ system with $5\%$ concentration is found to have the most stable ferromagnetic ground state. Its Curie temperature can achieve up to 173~K, which makes it a good candidate for spintronics applications.\\

\section{Acknowledgements}

The Authors thank Yuan Ping Feng, Chun Zhang, Jiajun Linhu for their helpful discussions and comments. The first-principles calculations were carried out on the GRC-NUS high-performance computing facilities. L.S. would like to acknowledge support from the ACRF Tier 1 Research Project (Project No. R-144-000-361-112). L.S. is member of the Singapore Spintronics Consortium (SG-SPIN). \\

\textbf{APPENDIX: BAND STRUCTURE AND PHONON DISPERSION}
\begin{figure}[ptb]
\centering
\includegraphics [width=0.60\textwidth]{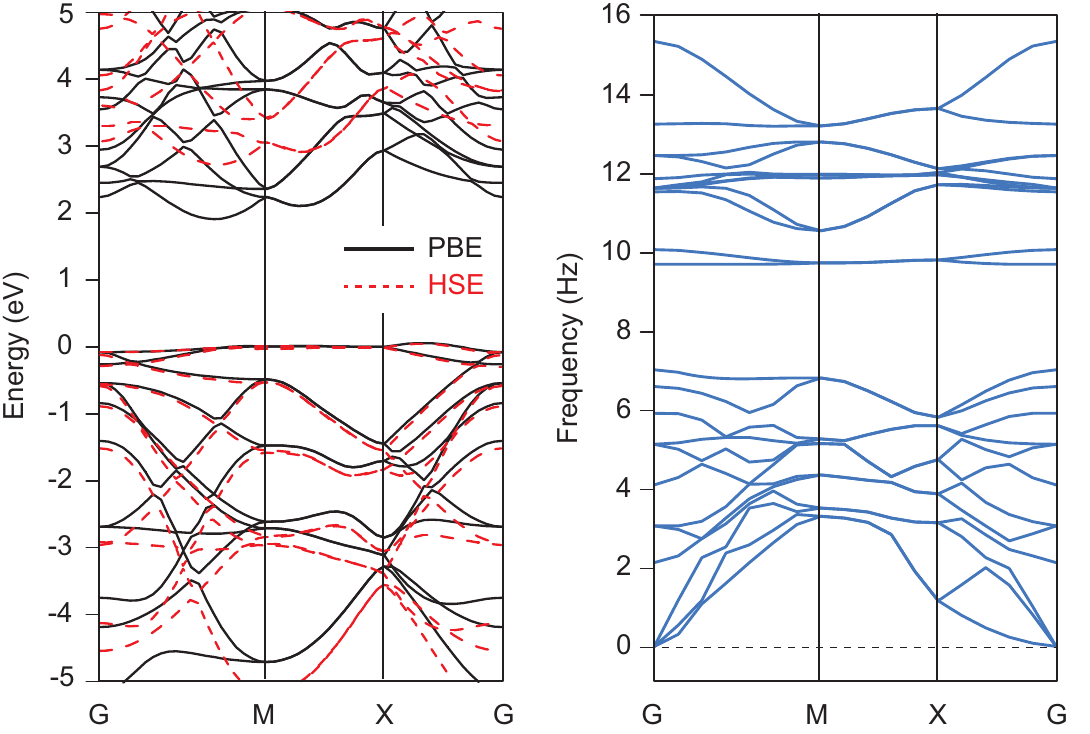}
\caption{{\footnotesize{} (a) The calculated band structure of $ot$-phosphorus. The band gap is 1.96 eV using PBE and 2.77 eV using HSE. (b) The phonon dispersion of $ot$-phosphorus. It shows that the structure of $ot$-phosphorus is stable since there is no non-trivial imaginary frequency. \label{fig:T-dependent-C}}}
\end{figure}

\centering

\newpage


\begin{thebibliography}{52}%
\makeatletter
\providecommand \@ifxundefined [1]{%
 \@ifx{#1\undefined}
}%
\providecommand \@ifnum [1]{%
 \ifnum #1\expandafter \@firstoftwo
 \else \expandafter \@secondoftwo
 \fi
}%
\providecommand \@ifx [1]{%
 \ifx #1\expandafter \@firstoftwo
 \else \expandafter \@secondoftwo
 \fi
}%
\providecommand \natexlab [1]{#1}%
\providecommand \enquote  [1]{``#1''}%
\providecommand \bibnamefont  [1]{#1}%
\providecommand \bibfnamefont [1]{#1}%
\providecommand \citenamefont [1]{#1}%
\providecommand \href@noop [0]{\@secondoftwo}%
\providecommand \href [0]{\begingroup \@sanitize@url \@href}%
\providecommand \@href[1]{\@@startlink{#1}\@@href}%
\providecommand \@@href[1]{\endgroup#1\@@endlink}%
\providecommand \@sanitize@url [0]{\catcode `\\12\catcode `\$12\catcode
  `\&12\catcode `\#12\catcode `\^12\catcode `\_12\catcode `\%12\relax}%
\providecommand \@@startlink[1]{}%
\providecommand \@@endlink[0]{}%
\providecommand \url  [0]{\begingroup\@sanitize@url \@url }%
\providecommand \@url [1]{\endgroup\@href {#1}{\urlprefix }}%
\providecommand \urlprefix  [0]{URL }%
\providecommand \Eprint [0]{\href }%
\providecommand \doibase [0]{http://dx.doi.org/}%
\providecommand \selectlanguage [0]{\@gobble}%
\providecommand \bibinfo  [0]{\@secondoftwo}%
\providecommand \bibfield  [0]{\@secondoftwo}%
\providecommand \translation [1]{[#1]}%
\providecommand \BibitemOpen [0]{}%
\providecommand \bibitemStop [0]{}%
\providecommand \bibitemNoStop [0]{.\EOS\space}%
\providecommand \EOS [0]{\spacefactor3000\relax}%
\providecommand \BibitemShut  [1]{\csname bibitem#1\endcsname}%
\let\auto@bib@innerbib\@empty
\bibitem [{\citenamefont {Butler}\ \emph {et~al.}(2013)\citenamefont {Butler},
  \citenamefont {Hollen}, \citenamefont {Cao}, \citenamefont {Cui},
  \citenamefont {Gupta}, \citenamefont {Guti{\'e}rrez}, \citenamefont {Heinz},
  \citenamefont {Hong}, \citenamefont {Huang}, \citenamefont {Ismach} \emph
  {et~al.}}]{butler2013progress}%
  \BibitemOpen
  \bibfield  {author} {\bibinfo {author} {\bibfnamefont {S.~Z.}\ \bibnamefont
  {Butler}}, \bibinfo {author} {\bibfnamefont {S.~M.}\ \bibnamefont {Hollen}},
  \bibinfo {author} {\bibfnamefont {L.}~\bibnamefont {Cao}}, \bibinfo {author}
  {\bibfnamefont {Y.}~\bibnamefont {Cui}}, \bibinfo {author} {\bibfnamefont
  {J.~A.}\ \bibnamefont {Gupta}}, \bibinfo {author} {\bibfnamefont {H.~R.}\
  \bibnamefont {Guti{\'e}rrez}}, \bibinfo {author} {\bibfnamefont {T.~F.}\
  \bibnamefont {Heinz}}, \bibinfo {author} {\bibfnamefont {S.~S.}\ \bibnamefont
  {Hong}}, \bibinfo {author} {\bibfnamefont {J.}~\bibnamefont {Huang}},
  \bibinfo {author} {\bibfnamefont {A.~F.}\ \bibnamefont {Ismach}},  \emph
  {et~al.},\ }\href@noop {} {\bibfield  {journal} {\bibinfo  {journal} {Acs
  Nano}\ }\textbf {\bibinfo {volume} {7}},\ \bibinfo {pages} {2898} (\bibinfo
  {year} {2013})}\BibitemShut {NoStop}%
\bibitem [{\citenamefont {Wang}\ \emph {et~al.}(2012)\citenamefont {Wang},
  \citenamefont {Kalantar-Zadeh}, \citenamefont {Kis}, \citenamefont
  {Coleman},\ and\ \citenamefont {Strano}}]{wang2012electronics}%
  \BibitemOpen
  \bibfield  {author} {\bibinfo {author} {\bibfnamefont {Q.~H.}\ \bibnamefont
  {Wang}}, \bibinfo {author} {\bibfnamefont {K.}~\bibnamefont
  {Kalantar-Zadeh}}, \bibinfo {author} {\bibfnamefont {A.}~\bibnamefont {Kis}},
  \bibinfo {author} {\bibfnamefont {J.~N.}\ \bibnamefont {Coleman}}, \ and\
  \bibinfo {author} {\bibfnamefont {M.~S.}\ \bibnamefont {Strano}},\
  }\href@noop {} {\bibfield  {journal} {\bibinfo  {journal} {Nat.
  Nanotechnol.}\ }\textbf {\bibinfo {volume} {7}},\ \bibinfo {pages} {699}
  (\bibinfo {year} {2012})}\BibitemShut {NoStop}%
\bibitem [{\citenamefont {Lopez-Bezanilla}\ \emph {et~al.}(2011)\citenamefont
  {Lopez-Bezanilla}, \citenamefont {Huang}, \citenamefont {Terrones},\ and\
  \citenamefont {Sumpter}}]{lopez2011boron}%
  \BibitemOpen
  \bibfield  {author} {\bibinfo {author} {\bibfnamefont {A.}~\bibnamefont
  {Lopez-Bezanilla}}, \bibinfo {author} {\bibfnamefont {J.}~\bibnamefont
  {Huang}}, \bibinfo {author} {\bibfnamefont {H.}~\bibnamefont {Terrones}}, \
  and\ \bibinfo {author} {\bibfnamefont {B.~G.}\ \bibnamefont {Sumpter}},\
  }\href@noop {} {\bibfield  {journal} {\bibinfo  {journal} {Nano. Lett.}\
  }\textbf {\bibinfo {volume} {11}},\ \bibinfo {pages} {3267} (\bibinfo {year}
  {2011})}\BibitemShut {NoStop}%
\bibitem [{\citenamefont {Reich}(2014)}]{Reich2014Nature}%
  \BibitemOpen
  \bibfield  {author} {\bibinfo {author} {\bibfnamefont {E.~S.}\ \bibnamefont
  {Reich}},\ }\href@noop {} {\bibfield  {journal} {\bibinfo  {journal}
  {Nature}\ }\textbf {\bibinfo {volume} {506}},\ \bibinfo {pages} {19}
  (\bibinfo {year} {2014})}\BibitemShut {NoStop}%
\bibitem [{\citenamefont {Li}\ \emph {et~al.}(2014)\citenamefont {Li},
  \citenamefont {Yu}, \citenamefont {Ye}, \citenamefont {Ge}, \citenamefont
  {Ou}, \citenamefont {Wu}, \citenamefont {Feng}, \citenamefont {Chen},\ and\
  \citenamefont {Zhang}}]{Li2014NN}%
  \BibitemOpen
  \bibfield  {author} {\bibinfo {author} {\bibfnamefont {L.}~\bibnamefont
  {Li}}, \bibinfo {author} {\bibfnamefont {Y.}~\bibnamefont {Yu}}, \bibinfo
  {author} {\bibfnamefont {G.~J.}\ \bibnamefont {Ye}}, \bibinfo {author}
  {\bibfnamefont {Q.}~\bibnamefont {Ge}}, \bibinfo {author} {\bibfnamefont
  {X.}~\bibnamefont {Ou}}, \bibinfo {author} {\bibfnamefont {H.}~\bibnamefont
  {Wu}}, \bibinfo {author} {\bibfnamefont {D.}~\bibnamefont {Feng}}, \bibinfo
  {author} {\bibfnamefont {X.~H.}\ \bibnamefont {Chen}}, \ and\ \bibinfo
  {author} {\bibfnamefont {Y.}~\bibnamefont {Zhang}},\ }\href@noop {}
  {\bibfield  {journal} {\bibinfo  {journal} {Nat. Nanotechnol.}\ }\textbf
  {\bibinfo {volume} {9}},\ \bibinfo {pages} {372} (\bibinfo {year}
  {2014})}\BibitemShut {NoStop}%
\bibitem [{\citenamefont {Liu}\ \emph {et~al.}(2014)\citenamefont {Liu},
  \citenamefont {Neal}, \citenamefont {Zhu}, \citenamefont {Luo}, \citenamefont
  {Xu}, \citenamefont {Tom{\'a}nek},\ and\ \citenamefont {Ye}}]{Liu2014AN}%
  \BibitemOpen
  \bibfield  {author} {\bibinfo {author} {\bibfnamefont {H.}~\bibnamefont
  {Liu}}, \bibinfo {author} {\bibfnamefont {A.~T.}\ \bibnamefont {Neal}},
  \bibinfo {author} {\bibfnamefont {Z.}~\bibnamefont {Zhu}}, \bibinfo {author}
  {\bibfnamefont {Z.}~\bibnamefont {Luo}}, \bibinfo {author} {\bibfnamefont
  {X.}~\bibnamefont {Xu}}, \bibinfo {author} {\bibfnamefont {D.}~\bibnamefont
  {Tom{\'a}nek}}, \ and\ \bibinfo {author} {\bibfnamefont {P.~D.}\ \bibnamefont
  {Ye}},\ }\href@noop {} {\bibfield  {journal} {\bibinfo  {journal} {Acs Nano}\
  } (\bibinfo {year} {2014})}\BibitemShut {NoStop}%
\bibitem [{\citenamefont {Koenig}\ \emph {et~al.}(2014)\citenamefont {Koenig},
  \citenamefont {Doganov}, \citenamefont {Schmidt}, \citenamefont {Neto},\ and\
  \citenamefont {Oezyilmaz}}]{Koenig2014APL}%
  \BibitemOpen
  \bibfield  {author} {\bibinfo {author} {\bibfnamefont {S.~P.}\ \bibnamefont
  {Koenig}}, \bibinfo {author} {\bibfnamefont {R.~A.}\ \bibnamefont {Doganov}},
  \bibinfo {author} {\bibfnamefont {H.}~\bibnamefont {Schmidt}}, \bibinfo
  {author} {\bibfnamefont {A.~C.}\ \bibnamefont {Neto}}, \ and\ \bibinfo
  {author} {\bibfnamefont {B.}~\bibnamefont {Oezyilmaz}},\ }\href@noop {}
  {\bibfield  {journal} {\bibinfo  {journal} {Appl. Phys. Lett.}\ }\textbf
  {\bibinfo {volume} {104}},\ \bibinfo {pages} {103106} (\bibinfo {year}
  {2014})}\BibitemShut {NoStop}%
\bibitem [{\citenamefont {Rodin}\ \emph {et~al.}(2014)\citenamefont {Rodin},
  \citenamefont {Carvalho},\ and\ \citenamefont {Castro~Neto}}]{Rodin2014PRL}%
  \BibitemOpen
  \bibfield  {author} {\bibinfo {author} {\bibfnamefont {A.~S.}\ \bibnamefont
  {Rodin}}, \bibinfo {author} {\bibfnamefont {A.}~\bibnamefont {Carvalho}}, \
  and\ \bibinfo {author} {\bibfnamefont {A.~H.}\ \bibnamefont {Castro~Neto}},\
  }\href {\doibase 10.1103/PhysRevLett.112.176801} {\bibfield  {journal}
  {\bibinfo  {journal} {Phys. Rev. Lett.}\ }\textbf {\bibinfo {volume} {112}},\
  \bibinfo {pages} {176801} (\bibinfo {year} {2014})}\BibitemShut {NoStop}%
\bibitem [{\citenamefont {Peng}\ \emph {et~al.}(2014)\citenamefont {Peng},
  \citenamefont {Wei},\ and\ \citenamefont {Copple}}]{Peng2014PRB}%
  \BibitemOpen
  \bibfield  {author} {\bibinfo {author} {\bibfnamefont {X.}~\bibnamefont
  {Peng}}, \bibinfo {author} {\bibfnamefont {Q.}~\bibnamefont {Wei}}, \ and\
  \bibinfo {author} {\bibfnamefont {A.}~\bibnamefont {Copple}},\ }\href
  {\doibase 10.1103/PhysRevB.90.085402} {\bibfield  {journal} {\bibinfo
  {journal} {Phys. Rev. B}\ }\textbf {\bibinfo {volume} {90}},\ \bibinfo
  {pages} {085402} (\bibinfo {year} {2014})}\BibitemShut {NoStop}%
\bibitem [{\citenamefont {Zhu}\ and\ \citenamefont
  {Tom\'anek}(2014)}]{Zhu2014PRL}%
  \BibitemOpen
  \bibfield  {author} {\bibinfo {author} {\bibfnamefont {Z.}~\bibnamefont
  {Zhu}}\ and\ \bibinfo {author} {\bibfnamefont {D.}~\bibnamefont
  {Tom\'anek}},\ }\href {\doibase 10.1103/PhysRevLett.112.176802} {\bibfield
  {journal} {\bibinfo  {journal} {Phys. Rev. Lett.}\ }\textbf {\bibinfo
  {volume} {112}},\ \bibinfo {pages} {176802} (\bibinfo {year}
  {2014})}\BibitemShut {NoStop}%
\bibitem [{\citenamefont {Zhu}\ \emph {et~al.}(2015)\citenamefont {Zhu},
  \citenamefont {Guan},\ and\ \citenamefont {Tom\'anek}}]{Zhu2015PRB}%
  \BibitemOpen
  \bibfield  {author} {\bibinfo {author} {\bibfnamefont {Z.}~\bibnamefont
  {Zhu}}, \bibinfo {author} {\bibfnamefont {J.}~\bibnamefont {Guan}}, \ and\
  \bibinfo {author} {\bibfnamefont {D.}~\bibnamefont {Tom\'anek}},\ }\href
  {\doibase 10.1103/PhysRevB.91.161404} {\bibfield  {journal} {\bibinfo
  {journal} {Phys. Rev. B}\ }\textbf {\bibinfo {volume} {91}},\ \bibinfo
  {pages} {161404} (\bibinfo {year} {2015})}\BibitemShut {NoStop}%
\bibitem [{\citenamefont {Guan}\ \emph
  {et~al.}(2014{\natexlab{a}})\citenamefont {Guan}, \citenamefont {Jin},
  \citenamefont {Zhu}, \citenamefont {Chuang}, \citenamefont {Jin},\ and\
  \citenamefont {Tom\'anek}}]{Guan2014PRB}%
  \BibitemOpen
  \bibfield  {author} {\bibinfo {author} {\bibfnamefont {J.}~\bibnamefont
  {Guan}}, \bibinfo {author} {\bibfnamefont {Z.}~\bibnamefont {Jin}}, \bibinfo
  {author} {\bibfnamefont {Z.}~\bibnamefont {Zhu}}, \bibinfo {author}
  {\bibfnamefont {C.}~\bibnamefont {Chuang}}, \bibinfo {author} {\bibfnamefont
  {B.-Y.}\ \bibnamefont {Jin}}, \ and\ \bibinfo {author} {\bibfnamefont
  {D.}~\bibnamefont {Tom\'anek}},\ }\href {\doibase 10.1103/PhysRevB.90.245403}
  {\bibfield  {journal} {\bibinfo  {journal} {Phys. Rev. B}\ }\textbf {\bibinfo
  {volume} {90}},\ \bibinfo {pages} {245403} (\bibinfo {year}
  {2014}{\natexlab{a}})}\BibitemShut {NoStop}%
\bibitem [{\citenamefont {Guan}\ \emph {et~al.}(2016)\citenamefont {Guan},
  \citenamefont {Song}, \citenamefont {Yang},\ and\ \citenamefont
  {Tom\'anek}}]{Guan2016PRB}%
  \BibitemOpen
  \bibfield  {author} {\bibinfo {author} {\bibfnamefont {J.}~\bibnamefont
  {Guan}}, \bibinfo {author} {\bibfnamefont {W.}~\bibnamefont {Song}}, \bibinfo
  {author} {\bibfnamefont {L.}~\bibnamefont {Yang}}, \ and\ \bibinfo {author}
  {\bibfnamefont {D.}~\bibnamefont {Tom\'anek}},\ }\href {\doibase
  10.1103/PhysRevB.94.045414} {\bibfield  {journal} {\bibinfo  {journal} {Phys.
  Rev. B}\ }\textbf {\bibinfo {volume} {94}},\ \bibinfo {pages} {045414}
  (\bibinfo {year} {2016})}\BibitemShut {NoStop}%
\bibitem [{\citenamefont {Fei}\ and\ \citenamefont {Yang}(2014)}]{Fei2014NL}%
  \BibitemOpen
  \bibfield  {author} {\bibinfo {author} {\bibfnamefont {R.}~\bibnamefont
  {Fei}}\ and\ \bibinfo {author} {\bibfnamefont {L.}~\bibnamefont {Yang}},\
  }\href@noop {} {\bibfield  {journal} {\bibinfo  {journal} {Nano. Lett.}\
  }\textbf {\bibinfo {volume} {14}},\ \bibinfo {pages} {2884} (\bibinfo {year}
  {2014})}\BibitemShut {NoStop}%
\bibitem [{\citenamefont {Fei}\ \emph {et~al.}(2015)\citenamefont {Fei},
  \citenamefont {Tran},\ and\ \citenamefont {Yang}}]{Fei2015PRB}%
  \BibitemOpen
  \bibfield  {author} {\bibinfo {author} {\bibfnamefont {R.}~\bibnamefont
  {Fei}}, \bibinfo {author} {\bibfnamefont {V.}~\bibnamefont {Tran}}, \ and\
  \bibinfo {author} {\bibfnamefont {L.}~\bibnamefont {Yang}},\ }\href {\doibase
  10.1103/PhysRevB.91.195319} {\bibfield  {journal} {\bibinfo  {journal} {Phys.
  Rev. B}\ }\textbf {\bibinfo {volume} {91}},\ \bibinfo {pages} {195319}
  (\bibinfo {year} {2015})}\BibitemShut {NoStop}%
\bibitem [{\citenamefont {Tran}\ \emph {et~al.}(2014)\citenamefont {Tran},
  \citenamefont {Soklaski}, \citenamefont {Liang},\ and\ \citenamefont
  {Yang}}]{Tran2014PRB}%
  \BibitemOpen
  \bibfield  {author} {\bibinfo {author} {\bibfnamefont {V.}~\bibnamefont
  {Tran}}, \bibinfo {author} {\bibfnamefont {R.}~\bibnamefont {Soklaski}},
  \bibinfo {author} {\bibfnamefont {Y.}~\bibnamefont {Liang}}, \ and\ \bibinfo
  {author} {\bibfnamefont {L.}~\bibnamefont {Yang}},\ }\href {\doibase
  10.1103/PhysRevB.89.235319} {\bibfield  {journal} {\bibinfo  {journal} {Phys.
  Rev. B}\ }\textbf {\bibinfo {volume} {89}},\ \bibinfo {pages} {235319}
  (\bibinfo {year} {2014})}\BibitemShut {NoStop}%
\bibitem [{\citenamefont {Tran}\ and\ \citenamefont
  {Yang}(2014)}]{Tran2014PRB2}%
  \BibitemOpen
  \bibfield  {author} {\bibinfo {author} {\bibfnamefont {V.}~\bibnamefont
  {Tran}}\ and\ \bibinfo {author} {\bibfnamefont {L.}~\bibnamefont {Yang}},\
  }\href {\doibase 10.1103/PhysRevB.89.245407} {\bibfield  {journal} {\bibinfo
  {journal} {Phys. Rev. B}\ }\textbf {\bibinfo {volume} {89}},\ \bibinfo
  {pages} {245407} (\bibinfo {year} {2014})}\BibitemShut {NoStop}%
\bibitem [{\citenamefont {Ghosh}\ \emph {et~al.}(2016)\citenamefont {Ghosh},
  \citenamefont {Singh}, \citenamefont {Prasad},\ and\ \citenamefont
  {Agarwal}}]{Ghosh2016PRB}%
  \BibitemOpen
  \bibfield  {author} {\bibinfo {author} {\bibfnamefont {B.}~\bibnamefont
  {Ghosh}}, \bibinfo {author} {\bibfnamefont {B.}~\bibnamefont {Singh}},
  \bibinfo {author} {\bibfnamefont {R.}~\bibnamefont {Prasad}}, \ and\ \bibinfo
  {author} {\bibfnamefont {A.}~\bibnamefont {Agarwal}},\ }\href {\doibase
  10.1103/PhysRevB.94.205426} {\bibfield  {journal} {\bibinfo  {journal} {Phys.
  Rev. B}\ }\textbf {\bibinfo {volume} {94}},\ \bibinfo {pages} {205426}
  (\bibinfo {year} {2016})}\BibitemShut {NoStop}%
\bibitem [{\citenamefont {Wu}\ \emph {et~al.}(2015)\citenamefont {Wu},
  \citenamefont {Shen}, \citenamefont {Yang}, \citenamefont {Cai},
  \citenamefont {Huang},\ and\ \citenamefont {Feng}}]{Wu2015PRB}%
  \BibitemOpen
  \bibfield  {author} {\bibinfo {author} {\bibfnamefont {Q.}~\bibnamefont
  {Wu}}, \bibinfo {author} {\bibfnamefont {L.}~\bibnamefont {Shen}}, \bibinfo
  {author} {\bibfnamefont {M.}~\bibnamefont {Yang}}, \bibinfo {author}
  {\bibfnamefont {Y.}~\bibnamefont {Cai}}, \bibinfo {author} {\bibfnamefont
  {Z.}~\bibnamefont {Huang}}, \ and\ \bibinfo {author} {\bibfnamefont {Y.~P.}\
  \bibnamefont {Feng}},\ }\href {\doibase 10.1103/PhysRevB.92.035436}
  {\bibfield  {journal} {\bibinfo  {journal} {Phys. Rev. B}\ }\textbf {\bibinfo
  {volume} {92}},\ \bibinfo {pages} {035436} (\bibinfo {year}
  {2015})}\BibitemShut {NoStop}%
\bibitem [{\citenamefont {Gomes}\ and\ \citenamefont
  {Carvalho}(2015)}]{Gomes2015PRB}%
  \BibitemOpen
  \bibfield  {author} {\bibinfo {author} {\bibfnamefont {L.~C.}\ \bibnamefont
  {Gomes}}\ and\ \bibinfo {author} {\bibfnamefont {A.}~\bibnamefont
  {Carvalho}},\ }\href {\doibase 10.1103/PhysRevB.92.085406} {\bibfield
  {journal} {\bibinfo  {journal} {Phys. Rev. B}\ }\textbf {\bibinfo {volume}
  {92}},\ \bibinfo {pages} {085406} (\bibinfo {year} {2015})}\BibitemShut
  {NoStop}%
\bibitem [{\citenamefont {Ziletti}\ \emph
  {et~al.}(2015{\natexlab{a}})\citenamefont {Ziletti}, \citenamefont
  {Carvalho}, \citenamefont {Trevisanutto}, \citenamefont {Campbell},
  \citenamefont {Coker},\ and\ \citenamefont {Castro~Neto}}]{Ziletti2015PRB}%
  \BibitemOpen
  \bibfield  {author} {\bibinfo {author} {\bibfnamefont {A.}~\bibnamefont
  {Ziletti}}, \bibinfo {author} {\bibfnamefont {A.}~\bibnamefont {Carvalho}},
  \bibinfo {author} {\bibfnamefont {P.~E.}\ \bibnamefont {Trevisanutto}},
  \bibinfo {author} {\bibfnamefont {D.~K.}\ \bibnamefont {Campbell}}, \bibinfo
  {author} {\bibfnamefont {D.~F.}\ \bibnamefont {Coker}}, \ and\ \bibinfo
  {author} {\bibfnamefont {A.~H.}\ \bibnamefont {Castro~Neto}},\ }\href
  {\doibase 10.1103/PhysRevB.91.085407} {\bibfield  {journal} {\bibinfo
  {journal} {Phys. Rev. B}\ }\textbf {\bibinfo {volume} {91}},\ \bibinfo
  {pages} {085407} (\bibinfo {year} {2015}{\natexlab{a}})}\BibitemShut
  {NoStop}%
\bibitem [{\citenamefont {Elahi}\ \emph {et~al.}(2015)\citenamefont {Elahi},
  \citenamefont {Khaliji}, \citenamefont {Tabatabaei}, \citenamefont
  {Pourfath},\ and\ \citenamefont {Asgari}}]{Elahi2015PRB}%
  \BibitemOpen
  \bibfield  {author} {\bibinfo {author} {\bibfnamefont {M.}~\bibnamefont
  {Elahi}}, \bibinfo {author} {\bibfnamefont {K.}~\bibnamefont {Khaliji}},
  \bibinfo {author} {\bibfnamefont {S.~M.}\ \bibnamefont {Tabatabaei}},
  \bibinfo {author} {\bibfnamefont {M.}~\bibnamefont {Pourfath}}, \ and\
  \bibinfo {author} {\bibfnamefont {R.}~\bibnamefont {Asgari}},\ }\href
  {\doibase 10.1103/PhysRevB.91.115412} {\bibfield  {journal} {\bibinfo
  {journal} {Phys. Rev. B}\ }\textbf {\bibinfo {volume} {91}},\ \bibinfo
  {pages} {115412} (\bibinfo {year} {2015})}\BibitemShut {NoStop}%
\bibitem [{\citenamefont {\ifmmode \mbox{\c{C}}\else \c{C}\fi{}ak\ifmmode
  \imath \else~\i \fi{}r}\ \emph {et~al.}(2015)\citenamefont {\ifmmode
  \mbox{\c{C}}\else \c{C}\fi{}ak\ifmmode \imath \else~\i \fi{}r}, \citenamefont
  {Sevik},\ and\ \citenamefont {Peeters}}]{Cakir2015PRB}%
  \BibitemOpen
  \bibfield  {author} {\bibinfo {author} {\bibfnamefont {D.}~\bibnamefont
  {\ifmmode \mbox{\c{C}}\else \c{C}\fi{}ak\ifmmode \imath \else~\i \fi{}r}},
  \bibinfo {author} {\bibfnamefont {C.}~\bibnamefont {Sevik}}, \ and\ \bibinfo
  {author} {\bibfnamefont {F.~M.}\ \bibnamefont {Peeters}},\ }\href {\doibase
  10.1103/PhysRevB.92.165406} {\bibfield  {journal} {\bibinfo  {journal} {Phys.
  Rev. B}\ }\textbf {\bibinfo {volume} {92}},\ \bibinfo {pages} {165406}
  (\bibinfo {year} {2015})}\BibitemShut {NoStop}%
\bibitem [{\citenamefont {Maity}\ \emph {et~al.}(2016)\citenamefont {Maity},
  \citenamefont {Singh}, \citenamefont {Sen}, \citenamefont {Kibey},
  \citenamefont {Kshirsagar},\ and\ \citenamefont {Kanhere}}]{Maity2016PRB}%
  \BibitemOpen
  \bibfield  {author} {\bibinfo {author} {\bibfnamefont {A.}~\bibnamefont
  {Maity}}, \bibinfo {author} {\bibfnamefont {A.}~\bibnamefont {Singh}},
  \bibinfo {author} {\bibfnamefont {P.}~\bibnamefont {Sen}}, \bibinfo {author}
  {\bibfnamefont {A.}~\bibnamefont {Kibey}}, \bibinfo {author} {\bibfnamefont
  {A.}~\bibnamefont {Kshirsagar}}, \ and\ \bibinfo {author} {\bibfnamefont
  {D.~G.}\ \bibnamefont {Kanhere}},\ }\href {\doibase
  10.1103/PhysRevB.94.075422} {\bibfield  {journal} {\bibinfo  {journal} {Phys.
  Rev. B}\ }\textbf {\bibinfo {volume} {94}},\ \bibinfo {pages} {075422}
  (\bibinfo {year} {2016})}\BibitemShut {NoStop}%
\bibitem [{\citenamefont {Ma}\ \emph {et~al.}(2016)\citenamefont {Ma},
  \citenamefont {Geng}, \citenamefont {Deng}, \citenamefont {Chen},
  \citenamefont {Sheng},\ and\ \citenamefont {Xing}}]{Ma2016PRB}%
  \BibitemOpen
  \bibfield  {author} {\bibinfo {author} {\bibfnamefont {R.}~\bibnamefont
  {Ma}}, \bibinfo {author} {\bibfnamefont {H.}~\bibnamefont {Geng}}, \bibinfo
  {author} {\bibfnamefont {W.~Y.}\ \bibnamefont {Deng}}, \bibinfo {author}
  {\bibfnamefont {M.~N.}\ \bibnamefont {Chen}}, \bibinfo {author}
  {\bibfnamefont {L.}~\bibnamefont {Sheng}}, \ and\ \bibinfo {author}
  {\bibfnamefont {D.~Y.}\ \bibnamefont {Xing}},\ }\href {\doibase
  10.1103/PhysRevB.94.125410} {\bibfield  {journal} {\bibinfo  {journal} {Phys.
  Rev. B}\ }\textbf {\bibinfo {volume} {94}},\ \bibinfo {pages} {125410}
  (\bibinfo {year} {2016})}\BibitemShut {NoStop}%
\bibitem [{\citenamefont {Rivero}\ \emph {et~al.}(2015)\citenamefont {Rivero},
  \citenamefont {Horvath}, \citenamefont {Zhu}, \citenamefont {Guan},
  \citenamefont {Tom\'anek},\ and\ \citenamefont
  {Barraza-Lopez}}]{Rivero2015PRB}%
  \BibitemOpen
  \bibfield  {author} {\bibinfo {author} {\bibfnamefont {P.}~\bibnamefont
  {Rivero}}, \bibinfo {author} {\bibfnamefont {C.~M.}\ \bibnamefont {Horvath}},
  \bibinfo {author} {\bibfnamefont {Z.}~\bibnamefont {Zhu}}, \bibinfo {author}
  {\bibfnamefont {J.}~\bibnamefont {Guan}}, \bibinfo {author} {\bibfnamefont
  {D.}~\bibnamefont {Tom\'anek}}, \ and\ \bibinfo {author} {\bibfnamefont
  {S.}~\bibnamefont {Barraza-Lopez}},\ }\href {\doibase
  10.1103/PhysRevB.91.115413} {\bibfield  {journal} {\bibinfo  {journal} {Phys.
  Rev. B}\ }\textbf {\bibinfo {volume} {91}},\ \bibinfo {pages} {115413}
  (\bibinfo {year} {2015})}\BibitemShut {NoStop}%
\bibitem [{\citenamefont {Yang}\ \emph {et~al.}(2016)\citenamefont {Yang},
  \citenamefont {Xu}, \citenamefont {Zhang}, \citenamefont {Ma},\ and\
  \citenamefont {Wu}}]{Yang2016PRB}%
  \BibitemOpen
  \bibfield  {author} {\bibinfo {author} {\bibfnamefont {G.}~\bibnamefont
  {Yang}}, \bibinfo {author} {\bibfnamefont {S.}~\bibnamefont {Xu}}, \bibinfo
  {author} {\bibfnamefont {W.}~\bibnamefont {Zhang}}, \bibinfo {author}
  {\bibfnamefont {T.}~\bibnamefont {Ma}}, \ and\ \bibinfo {author}
  {\bibfnamefont {C.}~\bibnamefont {Wu}},\ }\href {\doibase
  10.1103/PhysRevB.94.075106} {\bibfield  {journal} {\bibinfo  {journal} {Phys.
  Rev. B}\ }\textbf {\bibinfo {volume} {94}},\ \bibinfo {pages} {075106}
  (\bibinfo {year} {2016})}\BibitemShut {NoStop}%
\bibitem [{\citenamefont {Seixas}\ \emph {et~al.}(2015)\citenamefont {Seixas},
  \citenamefont {Carvalho},\ and\ \citenamefont {Castro~Neto}}]{Seixas2015PRB}%
  \BibitemOpen
  \bibfield  {author} {\bibinfo {author} {\bibfnamefont {L.}~\bibnamefont
  {Seixas}}, \bibinfo {author} {\bibfnamefont {A.}~\bibnamefont {Carvalho}}, \
  and\ \bibinfo {author} {\bibfnamefont {A.~H.}\ \bibnamefont {Castro~Neto}},\
  }\href {\doibase 10.1103/PhysRevB.91.155138} {\bibfield  {journal} {\bibinfo
  {journal} {Phys. Rev. B}\ }\textbf {\bibinfo {volume} {91}},\ \bibinfo
  {pages} {155138} (\bibinfo {year} {2015})}\BibitemShut {NoStop}%
\bibitem [{\citenamefont {Hanakata}\ \emph {et~al.}(2016)\citenamefont
  {Hanakata}, \citenamefont {Carvalho}, \citenamefont {Campbell},\ and\
  \citenamefont {Park}}]{Hanakata2016PRB}%
  \BibitemOpen
  \bibfield  {author} {\bibinfo {author} {\bibfnamefont {P.~Z.}\ \bibnamefont
  {Hanakata}}, \bibinfo {author} {\bibfnamefont {A.}~\bibnamefont {Carvalho}},
  \bibinfo {author} {\bibfnamefont {D.~K.}\ \bibnamefont {Campbell}}, \ and\
  \bibinfo {author} {\bibfnamefont {H.~S.}\ \bibnamefont {Park}},\ }\href
  {\doibase 10.1103/PhysRevB.94.035304} {\bibfield  {journal} {\bibinfo
  {journal} {Phys. Rev. B}\ }\textbf {\bibinfo {volume} {94}},\ \bibinfo
  {pages} {035304} (\bibinfo {year} {2016})}\BibitemShut {NoStop}%
\bibitem [{\citenamefont {Fu}\ \emph {et~al.}(2015)\citenamefont {Fu},
  \citenamefont {Wu}, \citenamefont {Gu}, \citenamefont {Wu},\ and\
  \citenamefont {Wu}}]{Fu2015PRB}%
  \BibitemOpen
  \bibfield  {author} {\bibinfo {author} {\bibfnamefont {H.-H.}\ \bibnamefont
  {Fu}}, \bibinfo {author} {\bibfnamefont {D.-D.}\ \bibnamefont {Wu}}, \bibinfo
  {author} {\bibfnamefont {L.}~\bibnamefont {Gu}}, \bibinfo {author}
  {\bibfnamefont {M.}~\bibnamefont {Wu}}, \ and\ \bibinfo {author}
  {\bibfnamefont {R.}~\bibnamefont {Wu}},\ }\href {\doibase
  10.1103/PhysRevB.92.045418} {\bibfield  {journal} {\bibinfo  {journal} {Phys.
  Rev. B}\ }\textbf {\bibinfo {volume} {92}},\ \bibinfo {pages} {045418}
  (\bibinfo {year} {2015})}\BibitemShut {NoStop}%
\bibitem [{\citenamefont {Tahir}\ \emph {et~al.}(2015)\citenamefont {Tahir},
  \citenamefont {Vasilopoulos},\ and\ \citenamefont {Peeters}}]{Tahir2015PRB}%
  \BibitemOpen
  \bibfield  {author} {\bibinfo {author} {\bibfnamefont {M.}~\bibnamefont
  {Tahir}}, \bibinfo {author} {\bibfnamefont {P.}~\bibnamefont {Vasilopoulos}},
  \ and\ \bibinfo {author} {\bibfnamefont {F.~M.}\ \bibnamefont {Peeters}},\
  }\href {\doibase 10.1103/PhysRevB.92.045420} {\bibfield  {journal} {\bibinfo
  {journal} {Phys. Rev. B}\ }\textbf {\bibinfo {volume} {92}},\ \bibinfo
  {pages} {045420} (\bibinfo {year} {2015})}\BibitemShut {NoStop}%
\bibitem [{\citenamefont {Jiang}\ \emph {et~al.}(2015)\citenamefont {Jiang},
  \citenamefont {Rold\'an}, \citenamefont {Guinea},\ and\ \citenamefont
  {Low}}]{Jiang2015PRB}%
  \BibitemOpen
  \bibfield  {author} {\bibinfo {author} {\bibfnamefont {Y.}~\bibnamefont
  {Jiang}}, \bibinfo {author} {\bibfnamefont {R.}~\bibnamefont {Rold\'an}},
  \bibinfo {author} {\bibfnamefont {F.}~\bibnamefont {Guinea}}, \ and\ \bibinfo
  {author} {\bibfnamefont {T.}~\bibnamefont {Low}},\ }\href {\doibase
  10.1103/PhysRevB.92.085408} {\bibfield  {journal} {\bibinfo  {journal} {Phys.
  Rev. B}\ }\textbf {\bibinfo {volume} {92}},\ \bibinfo {pages} {085408}
  (\bibinfo {year} {2015})}\BibitemShut {NoStop}%
\bibitem [{\citenamefont {Pereira}\ and\ \citenamefont
  {Katsnelson}(2015)}]{Pereira2015PRB}%
  \BibitemOpen
  \bibfield  {author} {\bibinfo {author} {\bibfnamefont {J.~M.}\ \bibnamefont
  {Pereira}}\ and\ \bibinfo {author} {\bibfnamefont {M.~I.}\ \bibnamefont
  {Katsnelson}},\ }\href {\doibase 10.1103/PhysRevB.92.075437} {\bibfield
  {journal} {\bibinfo  {journal} {Phys. Rev. B}\ }\textbf {\bibinfo {volume}
  {92}},\ \bibinfo {pages} {075437} (\bibinfo {year} {2015})}\BibitemShut
  {NoStop}%
\bibitem [{\citenamefont {Ostahie}\ and\ \citenamefont
  {Aldea}(2016)}]{Ostahie2016PRB}%
  \BibitemOpen
  \bibfield  {author} {\bibinfo {author} {\bibfnamefont {B.}~\bibnamefont
  {Ostahie}}\ and\ \bibinfo {author} {\bibfnamefont {A.}~\bibnamefont
  {Aldea}},\ }\href {\doibase 10.1103/PhysRevB.93.075408} {\bibfield  {journal}
  {\bibinfo  {journal} {Phys. Rev. B}\ }\textbf {\bibinfo {volume} {93}},\
  \bibinfo {pages} {075408} (\bibinfo {year} {2016})}\BibitemShut {NoStop}%
\bibitem [{\citenamefont {Zhou}\ \emph {et~al.}(2015)\citenamefont {Zhou},
  \citenamefont {Lou}, \citenamefont {Zhai},\ and\ \citenamefont
  {Chang}}]{Zhou2015PRB}%
  \BibitemOpen
  \bibfield  {author} {\bibinfo {author} {\bibfnamefont {X.}~\bibnamefont
  {Zhou}}, \bibinfo {author} {\bibfnamefont {W.-K.}\ \bibnamefont {Lou}},
  \bibinfo {author} {\bibfnamefont {F.}~\bibnamefont {Zhai}}, \ and\ \bibinfo
  {author} {\bibfnamefont {K.}~\bibnamefont {Chang}},\ }\href {\doibase
  10.1103/PhysRevB.92.165405} {\bibfield  {journal} {\bibinfo  {journal} {Phys.
  Rev. B}\ }\textbf {\bibinfo {volume} {92}},\ \bibinfo {pages} {165405}
  (\bibinfo {year} {2015})}\BibitemShut {NoStop}%
\bibitem [{\citenamefont {Ziletti}\ \emph
  {et~al.}(2015{\natexlab{b}})\citenamefont {Ziletti}, \citenamefont
  {Carvalho}, \citenamefont {Campbell}, \citenamefont {Coker},\ and\
  \citenamefont {Castro~Neto}}]{Ziletti2015PRL}%
  \BibitemOpen
  \bibfield  {author} {\bibinfo {author} {\bibfnamefont {A.}~\bibnamefont
  {Ziletti}}, \bibinfo {author} {\bibfnamefont {A.}~\bibnamefont {Carvalho}},
  \bibinfo {author} {\bibfnamefont {D.~K.}\ \bibnamefont {Campbell}}, \bibinfo
  {author} {\bibfnamefont {D.~F.}\ \bibnamefont {Coker}}, \ and\ \bibinfo
  {author} {\bibfnamefont {A.~H.}\ \bibnamefont {Castro~Neto}},\ }\href
  {\doibase 10.1103/PhysRevLett.114.046801} {\bibfield  {journal} {\bibinfo
  {journal} {Phys. Rev. Lett.}\ }\textbf {\bibinfo {volume} {114}},\ \bibinfo
  {pages} {046801} (\bibinfo {year} {2015}{\natexlab{b}})}\BibitemShut
  {NoStop}%
\bibitem [{\citenamefont {Guan}\ \emph
  {et~al.}(2014{\natexlab{b}})\citenamefont {Guan}, \citenamefont {Zhu},\ and\
  \citenamefont {Tom\'anek}}]{Guan2014PRL}%
  \BibitemOpen
  \bibfield  {author} {\bibinfo {author} {\bibfnamefont {J.}~\bibnamefont
  {Guan}}, \bibinfo {author} {\bibfnamefont {Z.}~\bibnamefont {Zhu}}, \ and\
  \bibinfo {author} {\bibfnamefont {D.}~\bibnamefont {Tom\'anek}},\ }\href
  {\doibase 10.1103/PhysRevLett.113.226801} {\bibfield  {journal} {\bibinfo
  {journal} {Phys. Rev. Lett.}\ }\textbf {\bibinfo {volume} {113}},\ \bibinfo
  {pages} {226801} (\bibinfo {year} {2014}{\natexlab{b}})}\BibitemShut
  {NoStop}%
\bibitem [{\citenamefont {Guan}\ \emph
  {et~al.}(2014{\natexlab{c}})\citenamefont {Guan}, \citenamefont {Zhu},\ and\
  \citenamefont {Tom\'anek}}]{Guan2014PRL2}%
  \BibitemOpen
  \bibfield  {author} {\bibinfo {author} {\bibfnamefont {J.}~\bibnamefont
  {Guan}}, \bibinfo {author} {\bibfnamefont {Z.}~\bibnamefont {Zhu}}, \ and\
  \bibinfo {author} {\bibfnamefont {D.}~\bibnamefont {Tom\'anek}},\ }\href
  {\doibase 10.1103/PhysRevLett.113.046804} {\bibfield  {journal} {\bibinfo
  {journal} {Phys. Rev. Lett.}\ }\textbf {\bibinfo {volume} {113}},\ \bibinfo
  {pages} {046804} (\bibinfo {year} {2014}{\natexlab{c}})}\BibitemShut
  {NoStop}%
\bibitem [{\citenamefont {Boulfelfel}\ \emph {et~al.}(2012)\citenamefont
  {Boulfelfel}, \citenamefont {Seifert}, \citenamefont {Grin},\ and\
  \citenamefont {Leoni}}]{Boulfelfel2012PRB}%
  \BibitemOpen
  \bibfield  {author} {\bibinfo {author} {\bibfnamefont {S.~E.}\ \bibnamefont
  {Boulfelfel}}, \bibinfo {author} {\bibfnamefont {G.}~\bibnamefont {Seifert}},
  \bibinfo {author} {\bibfnamefont {Y.}~\bibnamefont {Grin}}, \ and\ \bibinfo
  {author} {\bibfnamefont {S.}~\bibnamefont {Leoni}},\ }\href {\doibase
  10.1103/PhysRevB.85.014110} {\bibfield  {journal} {\bibinfo  {journal} {Phys.
  Rev. B}\ }\textbf {\bibinfo {volume} {85}},\ \bibinfo {pages} {014110}
  (\bibinfo {year} {2012})}\BibitemShut {NoStop}%
\bibitem [{\citenamefont {Zhang}\ \emph {et~al.}(2015)\citenamefont {Zhang},
  \citenamefont {Lee}, \citenamefont {Wang},\ and\ \citenamefont
  {Yao}}]{Zhang2015CMS}%
  \BibitemOpen
  \bibfield  {author} {\bibinfo {author} {\bibfnamefont {Y.}~\bibnamefont
  {Zhang}}, \bibinfo {author} {\bibfnamefont {J.}~\bibnamefont {Lee}}, \bibinfo
  {author} {\bibfnamefont {W.-L.}\ \bibnamefont {Wang}}, \ and\ \bibinfo
  {author} {\bibfnamefont {D.-X.}\ \bibnamefont {Yao}},\ }\href@noop {}
  {\bibfield  {journal} {\bibinfo  {journal} {Comp. Mater. Sci.}\ }\textbf
  {\bibinfo {volume} {110}},\ \bibinfo {pages} {109} (\bibinfo {year}
  {2015})}\BibitemShut {NoStop}%
\bibitem [{\citenamefont {Zhao}\ \emph {et~al.}(2016)\citenamefont {Zhao},
  \citenamefont {Wang},\ and\ \citenamefont {Liao}}]{Zhao2016CPL}%
  \BibitemOpen
  \bibfield  {author} {\bibinfo {author} {\bibfnamefont {T.}~\bibnamefont
  {Zhao}}, \bibinfo {author} {\bibfnamefont {G.}~\bibnamefont {Wang}}, \ and\
  \bibinfo {author} {\bibfnamefont {Y.}~\bibnamefont {Liao}},\ }\href@noop {}
  {\bibfield  {journal} {\bibinfo  {journal} {Chem Phys Lett}\ }\textbf
  {\bibinfo {volume} {652}},\ \bibinfo {pages} {1} (\bibinfo {year}
  {2016})}\BibitemShut {NoStop}%
\bibitem [{\citenamefont {Sato}\ \emph {et~al.}(2010)\citenamefont {Sato},
  \citenamefont {Bergqvist}, \citenamefont {Kudrnovsk{\`y}}, \citenamefont
  {Dederichs}, \citenamefont {Eriksson}, \citenamefont {Turek}, \citenamefont
  {Sanyal}, \citenamefont {Bouzerar}, \citenamefont {Katayama-Yoshida},
  \citenamefont {Dinh} \emph {et~al.}}]{sato2010first}%
  \BibitemOpen
  \bibfield  {author} {\bibinfo {author} {\bibfnamefont {K.}~\bibnamefont
  {Sato}}, \bibinfo {author} {\bibfnamefont {L.}~\bibnamefont {Bergqvist}},
  \bibinfo {author} {\bibfnamefont {J.}~\bibnamefont {Kudrnovsk{\`y}}},
  \bibinfo {author} {\bibfnamefont {P.~H.}\ \bibnamefont {Dederichs}}, \bibinfo
  {author} {\bibfnamefont {O.}~\bibnamefont {Eriksson}}, \bibinfo {author}
  {\bibfnamefont {I.}~\bibnamefont {Turek}}, \bibinfo {author} {\bibfnamefont
  {B.}~\bibnamefont {Sanyal}}, \bibinfo {author} {\bibfnamefont
  {G.}~\bibnamefont {Bouzerar}}, \bibinfo {author} {\bibfnamefont
  {H.}~\bibnamefont {Katayama-Yoshida}}, \bibinfo {author} {\bibfnamefont
  {V.}~\bibnamefont {Dinh}},  \emph {et~al.},\ }\href@noop {} {\bibfield
  {journal} {\bibinfo  {journal} {Rev Mod Phys}\ }\textbf {\bibinfo {volume}
  {82}},\ \bibinfo {pages} {1633} (\bibinfo {year} {2010})}\BibitemShut
  {NoStop}%
\bibitem [{\citenamefont {Zhou}\ \emph {et~al.}(2013)\citenamefont {Zhou},
  \citenamefont {Zou}, \citenamefont {Najmaei}, \citenamefont {Liu},
  \citenamefont {Shi}, \citenamefont {Kong}, \citenamefont {Lou}, \citenamefont
  {Ajayan}, \citenamefont {Yakobson},\ and\ \citenamefont
  {Idrobo}}]{zhou2013intrinsic}%
  \BibitemOpen
  \bibfield  {author} {\bibinfo {author} {\bibfnamefont {W.}~\bibnamefont
  {Zhou}}, \bibinfo {author} {\bibfnamefont {X.}~\bibnamefont {Zou}}, \bibinfo
  {author} {\bibfnamefont {S.}~\bibnamefont {Najmaei}}, \bibinfo {author}
  {\bibfnamefont {Z.}~\bibnamefont {Liu}}, \bibinfo {author} {\bibfnamefont
  {Y.}~\bibnamefont {Shi}}, \bibinfo {author} {\bibfnamefont {J.}~\bibnamefont
  {Kong}}, \bibinfo {author} {\bibfnamefont {J.}~\bibnamefont {Lou}}, \bibinfo
  {author} {\bibfnamefont {P.~M.}\ \bibnamefont {Ajayan}}, \bibinfo {author}
  {\bibfnamefont {B.~I.}\ \bibnamefont {Yakobson}}, \ and\ \bibinfo {author}
  {\bibfnamefont {J.-C.}\ \bibnamefont {Idrobo}},\ }\href@noop {} {\bibfield
  {journal} {\bibinfo  {journal} {Nano. Lett.}\ }\textbf {\bibinfo {volume}
  {13}},\ \bibinfo {pages} {2615} (\bibinfo {year} {2013})}\BibitemShut
  {NoStop}%
\bibitem [{\citenamefont {Hohenberg}\ and\ \citenamefont
  {Kohn}(1964)}]{hohenberg1964inhomogeneous}%
  \BibitemOpen
  \bibfield  {author} {\bibinfo {author} {\bibfnamefont {P.}~\bibnamefont
  {Hohenberg}}\ and\ \bibinfo {author} {\bibfnamefont {W.}~\bibnamefont
  {Kohn}},\ }\href@noop {} {\bibfield  {journal} {\bibinfo  {journal} {Phys.
  Rev.}\ }\textbf {\bibinfo {volume} {136}},\ \bibinfo {pages} {B864} (\bibinfo
  {year} {1964})}\BibitemShut {NoStop}%
\bibitem [{\citenamefont {Kohn}\ and\ \citenamefont
  {Sham}(1965)}]{kohn1965self}%
  \BibitemOpen
  \bibfield  {author} {\bibinfo {author} {\bibfnamefont {W.}~\bibnamefont
  {Kohn}}\ and\ \bibinfo {author} {\bibfnamefont {L.~J.}\ \bibnamefont
  {Sham}},\ }\href@noop {} {\bibfield  {journal} {\bibinfo  {journal} {Phys.
  Rev.}\ }\textbf {\bibinfo {volume} {140}},\ \bibinfo {pages} {A1133}
  (\bibinfo {year} {1965})}\BibitemShut {NoStop}%
\bibitem [{\citenamefont {Kresse}\ and\ \citenamefont
  {Furthm{\"u}ller}(1996{\natexlab{a}})}]{Kresse1996PRB}%
  \BibitemOpen
  \bibfield  {author} {\bibinfo {author} {\bibfnamefont {G.}~\bibnamefont
  {Kresse}}\ and\ \bibinfo {author} {\bibfnamefont {J.}~\bibnamefont
  {Furthm{\"u}ller}},\ }\href@noop {} {\bibfield  {journal} {\bibinfo
  {journal} {Phys. Rev. B}\ }\textbf {\bibinfo {volume} {54}},\ \bibinfo
  {pages} {11169} (\bibinfo {year} {1996}{\natexlab{a}})}\BibitemShut {NoStop}%
\bibitem [{\citenamefont {Kresse}\ and\ \citenamefont
  {Joubert}(1999)}]{Kresse1999PRB}%
  \BibitemOpen
  \bibfield  {author} {\bibinfo {author} {\bibfnamefont {G.}~\bibnamefont
  {Kresse}}\ and\ \bibinfo {author} {\bibfnamefont {D.}~\bibnamefont
  {Joubert}},\ }\href@noop {} {\bibfield  {journal} {\bibinfo  {journal} {Phys.
  Rev. B}\ }\textbf {\bibinfo {volume} {59}},\ \bibinfo {pages} {1758}
  (\bibinfo {year} {1999})}\BibitemShut {NoStop}%
\bibitem [{\citenamefont {Kresse}\ and\ \citenamefont
  {Furthm{\"u}ller}(1996{\natexlab{b}})}]{vasp2}%
  \BibitemOpen
  \bibfield  {author} {\bibinfo {author} {\bibfnamefont {G.}~\bibnamefont
  {Kresse}}\ and\ \bibinfo {author} {\bibfnamefont {J.}~\bibnamefont
  {Furthm{\"u}ller}},\ }\href@noop {} {\bibfield  {journal} {\bibinfo
  {journal} {Comput. Mater. Sci.}\ }\textbf {\bibinfo {volume} {6}},\ \bibinfo
  {pages} {15} (\bibinfo {year} {1996}{\natexlab{b}})}\BibitemShut {NoStop}%
\bibitem [{\citenamefont {Bl{\"o}chl}(1994)}]{blochl1994projector}%
  \BibitemOpen
  \bibfield  {author} {\bibinfo {author} {\bibfnamefont {P.~E.}\ \bibnamefont
  {Bl{\"o}chl}},\ }\href@noop {} {\bibfield  {journal} {\bibinfo  {journal}
  {Phys. Rev. B}\ }\textbf {\bibinfo {volume} {50}},\ \bibinfo {pages} {17953}
  (\bibinfo {year} {1994})}\BibitemShut {NoStop}%
\bibitem [{\citenamefont {Dietl}(2010)}]{Dietl2010NM}%
  \BibitemOpen
  \bibfield  {author} {\bibinfo {author} {\bibfnamefont {T.}~\bibnamefont
  {Dietl}},\ }\href@noop {} {\bibfield  {journal} {\bibinfo  {journal} {Nat.
  Mater.}\ }\textbf {\bibinfo {volume} {9}},\ \bibinfo {pages} {965} (\bibinfo
  {year} {2010})}\BibitemShut {NoStop}%
\bibitem [{\citenamefont {Pan}\ \emph {et~al.}(2007)\citenamefont {Pan},
  \citenamefont {Yi}, \citenamefont {Shen}, \citenamefont {Wu}, \citenamefont
  {Yang}, \citenamefont {Lin}, \citenamefont {Feng}, \citenamefont {Ding},
  \citenamefont {Van},\ and\ \citenamefont {Yin}}]{Pan2007PRL}%
  \BibitemOpen
  \bibfield  {author} {\bibinfo {author} {\bibfnamefont {H.}~\bibnamefont
  {Pan}}, \bibinfo {author} {\bibfnamefont {J.~B.}\ \bibnamefont {Yi}},
  \bibinfo {author} {\bibfnamefont {L.}~\bibnamefont {Shen}}, \bibinfo {author}
  {\bibfnamefont {R.~Q.}\ \bibnamefont {Wu}}, \bibinfo {author} {\bibfnamefont
  {J.~H.}\ \bibnamefont {Yang}}, \bibinfo {author} {\bibfnamefont {J.~Y.}\
  \bibnamefont {Lin}}, \bibinfo {author} {\bibfnamefont {Y.~P.}\ \bibnamefont
  {Feng}}, \bibinfo {author} {\bibfnamefont {J.}~\bibnamefont {Ding}}, \bibinfo
  {author} {\bibfnamefont {L.~H.}\ \bibnamefont {Van}}, \ and\ \bibinfo
  {author} {\bibfnamefont {J.~H.}\ \bibnamefont {Yin}},\ }\href {\doibase
  10.1103/PhysRevLett.99.127201} {\bibfield  {journal} {\bibinfo  {journal}
  {Phys. Rev. Lett.}\ }\textbf {\bibinfo {volume} {99}},\ \bibinfo {pages}
  {127201} (\bibinfo {year} {2007})}\BibitemShut {NoStop}%
\bibitem [{\citenamefont {Sato}\ \emph {et~al.}(2003)\citenamefont {Sato},
  \citenamefont {Dederics},\ and\ \citenamefont
  {Katayama-Yoshida}}]{Sato2003EPL}%
  \BibitemOpen
  \bibfield  {author} {\bibinfo {author} {\bibfnamefont {K.}~\bibnamefont
  {Sato}}, \bibinfo {author} {\bibfnamefont {P.}~\bibnamefont {Dederics}}, \
  and\ \bibinfo {author} {\bibfnamefont {H.}~\bibnamefont {Katayama-Yoshida}},\
  }\href@noop {} {\bibfield  {journal} {\bibinfo  {journal} {EPL (Europhysics
  Letters)}\ }\textbf {\bibinfo {volume} {61}},\ \bibinfo {pages} {403}
  (\bibinfo {year} {2003})}\BibitemShut {NoStop}%
\end{thebibliography}
%

\end{document}